\documentclass[aps,pre,twocolumn,nofootinbib, longbibliography,preprint,10pt,unsortedaddress,titlepage]{revtex4-1}

\usepackage{graphicx}
\usepackage{dcolumn}
\usepackage{bm}

\usepackage[utf8]{inputenc}
\usepackage{mathtools}
\usepackage[T1]{fontenc}
\usepackage{mathptmx}
\usepackage[utf8]{inputenc}
\usepackage{graphicx}
\usepackage{subcaption}
\usepackage{amsmath}
\usepackage{amssymb}
\usepackage{dcolumn}
\usepackage{bm}
\usepackage{xr}
\usepackage{makecell}
\usepackage{cleveref}

\begin{document}

\title{Optimal Entanglement of Polymers Promotes Formation of Highly Oriented Fibers}
\author{Artem Petrov $^{\ast}$ \textit{$^{\dag}$}}
	\affiliation{Faculty of Physics, Lomonosov Moscow State University, 119991 Moscow, Russia}
	\author{Vladimir Yu. Rudyak \textit{$^{\dag}$}}
	\affiliation{Faculty of Physics, Lomonosov Moscow State University, 119991 Moscow, Russia}
	\author{Alexander Chertovich}
	\affiliation{Semenov Federal Research Center for Chemical Physics, 119991 Moscow, Russia}
	\affiliation{Faculty of Physics, Lomonosov Moscow State University, 119991 Moscow, Russia}
	\date{\today\clearpage}

\footnotetext{$\ast$~petrov.ai15@physics.msu.ru}
\footnotetext{\dag~These authors contributed equally to this work.}

\begin{abstract}
Polymer fibers consist of macromolecules oriented along the fiber axis. Better alignment of chains leads to an increased strength of the fiber. It is believed that the key factor preventing formation of a perfectly oriented fiber is entanglement of polymers. We performed large-scale computer simulations of uniaxial stretching of semicrystalline ultrahigh molecular weight polyethylene. We discovered that there is an optimal number of entanglements per macromolecule necessary to maximize chain orientation in a fiber. Polymers that were entangled too strongly formed less oriented fibers. On the other hand, when polymers had too few entanglements per chain, they disentangled during stretching, and the strong fiber was not formed. We constructed a microscopic analytical theory describing both the fiber formation and disentanglement processes. Our work presents a novel view on the role of entanglements during fiber production and predicts the existence of a single universal optimal number of entanglements per chain maximizing the fiber quality: approximately $10^2$ entanglements.
\end{abstract}

\maketitle

\section*{Introduction}

Synthetic polymer fibers such as nylon, polyester, and polyethylene are one of the main high-tonnage products in chemical industry. Alignment of polymer chains along the fiber axis is one of the simplest and most reliable ways to improve its mechanical properties \cite{carothers1932studies}. In the middle of the 20th century, the Young's modulus of an ideally oriented polyethylene (PE) fiber was predicted to be two orders of magnitude higher than for an isotropic PE sample \cite{TRELOAR196095,lyons1958theoretical}. Improvement of orientation of chain segments is the key to obtain stronger fibers.

Methods of obtaining oriented PE fibers include solution(gel)-spinning \cite{smith1980ultra1,smith1980ultra2}, electrospinning \cite{bhardwaj2010electrospinning,doshi1995electrospinning}, and the recently developed solvent-free techniques \cite{Rastogi11,rastogi2011unprecedented}. The popular hypothesis suggests that the presence of interchain entanglements, which appear due to uncrossablility of chains, limits stretching of a polymer sample \cite{rastogi2011unprecedented,smith1980ultra2,myasnikova2011reactor,Pennings83,christakopoulos2021melting}. Different techniques decrease the density of entanglements in different ways. During solution(gel)-spinning, for example, the entanglement density is reduced by placing the low concentrated polymer material in a good solvent \cite{smith1980ultra2}. Other works showed the possibility to decrease entanglement density by tuning the conditions of polymerization procedure \cite{Rastogi11,rastogi2011unprecedented,sano2001ultradrawing,gote2018judicious}. However, it is worth mentioning that the conclusions about the decrease of entanglement density are always based on indirect observations due to impossibility of its straightforward measurement.

After obtaining a polymer material with weakly entangled chains, the sample is drawn (usually uniaxially stretched) \cite{myasnikova2011reactor,rastogi2011unprecedented,smith1980ultra2,smith1982tensile}. It is during this stage when the process of orientation mainly occurs. Stretching is usually carried out below the crystallization temperature of a drawn polymer to prevent re-entanglement of disentangled chains \cite{myasnikova2011reactor,tian2014lamellae,tian2015transition}. Since complete crystallization is impossible in polymer systems, the stretched materials are in the so-called semicrystalline state consisting of crystallites and amorphous regions.

The microscopic mechanisms of tensile deformation of semicrystalline polymers were extensively studied in the last few decades. One of the most popular models describes tensile stretching as the deformation of interpenetrating amorphous and crystalline networks \cite{hiss1999network,hobeika2000temperature,men2003role,men2020critical}. Crystalline domains play critical role at the beginning of stretching (the elastic deformation) and determine the Young's modulus of a semicrystalline sample \cite{men2003role,men2020critical}. At larger strains, the elastic deformation regime ends, and the yield point and strain softening are observed due to partial break of crystallites. Deformation of the amorphous network is believed to be the dominant microscopic process after the yield point \cite{men2020critical,men2003role}. Further stretching causes the onset of strain hardening, during which the formation of oriented fibrils occurs. The so-called strain hardening modulus, which characterizes malleability of a sample during this stage of deformation, is proposed to be proportional to the density of entanglements situated in the amorphous phase \cite{men2003role}. In the samples of high molecular weight polymers, strain hardening proceeds sometimes via formation of the so-called "shish-kebab" structure \cite{tian2014lamellae,tian2015transition,Pennings83,myasnikova2011reactor}. The limiting strain is also believed to depend on the entanglement density \cite{haward1968use,arruda1993three,myasnikova2011reactor,fukuoka2006role}; however, direct experimental evidence for this is lacking. Summing up, there are no quantitative relations between the entanglement density and the characteristics of the drawn fiber.

Experimental works are limited in their ability to provide insight into the microscopic origins of the deformation behavior. As a result, various computer simulation approaches were developed to complement the existing experimental knowledge \cite{queyroy2012effect,monasse2008molecular,jabbari2015plastic,jabbari2017role,lee2011plastic,kim2014plastic,yeh2015mechanical,yeh2017molecular,schurmann1998polyethylene,Rutledge20,Higuchi19,jabbari2015correlation}. Several papers describe tensile stretching of a semicrystalline polymer sample as deformation of a "stack" consisting of crystalline and amorphous phases \cite{queyroy2012effect,monasse2008molecular,lee2011plastic,kim2014plastic,yeh2015mechanical,yeh2017molecular,Rutledge20,Higuchi19}. Treating the process of semicrystalline polymer deformation as stretching of a layered "stack" is a great simplification. A more realistic model of deformation was described in refs. \cite{jabbari2015plastic,jabbari2017role,jabbari2015correlation}. In those works, the authors carried out the deformation simulations using the coarse-grained model of semicrystalline poly(vinyl alcohol) (PVA) and PE described in ref. \cite{meyer2002formation}.

The aforementioned simulation approaches showed that the mechanism of deformation depends strongly on the deformation rate and on volume conservation. In particular, the authors found that polymers yield via formation of cavities except for the case of slow volume-conserving deformation \cite{lee2011plastic}. Usually cavitation is undesired during fiber formation, since it may cause early fracture. The choice of slow volume-conserving deformation technique is most reasonable for simulation of fiber drawing and a subsequent comparison with experimental data.

Despite the possibility of studying polymer entanglement directly in simulations \cite{everaers2004rheology,kroger2005shortest}, the effect of entanglement density on the process of polymer material deformation was studied very poorly. To the best of our knowledge, only one simulation study compared the mechanism of deformation in the samples with different densities of entanglements \cite{Higuchi19}. It has been shown that an increase of entanglement density delayed cavitation and decreased the stress at fracture point. The effect of entanglement density on the degree of fiber orientation was not considered in that paper. Hence, the influence of the entanglement density on the mechanical and structural properties of polymer fibers was not studied directly neither in simulations nor in the experiments.

Another long-standing problem in that field is the development of a corresponding analytical model. The earliest theories were designed in 1960s; however, the developed models of deformation were mostly semi-empirical \cite{haward1968use,g1979determination,lee1993simulation}. One of the main goals of those theories was to predict the stress-strain curves, i.e., the macroscopic response of a semicrystalline material on deformation. The microscopic models describing the evolution of chain folding in the stretched material are still lacking.

\begin{figure}
\centering
  \includegraphics[width=0.9\linewidth]{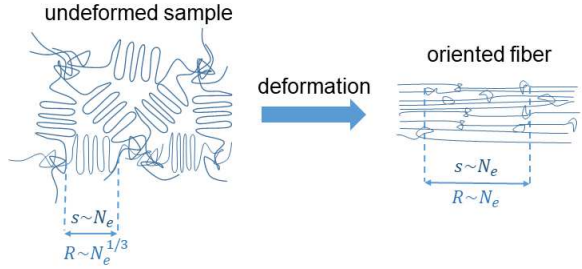}
  \caption{Schematic representation of fiber formation. The entanglement length $N_e$ is the number of monomer units between two consecutive entanglements along the chain. During deformation, the chain segments of length $s\approx N_e$ unfold and form straight segments that have larger end-to-end distance $R$.}
  \label{fig:1}
\end{figure}

In this work, we performed large-scale coarse-grained computer simulations of tensile deformation of semicrystalline polymers into oriented fibers, see sketch in Fig. \ref{fig:1}. We studied how stress-strain curves depended on the initial entanglement density in the sample. We also investigated the microscopic mechanism of fiber formation and explained our findings by an analytical model. We found out that the number of entanglements per chain should be optimal in order to obtain a strong highly oriented fiber. We derived this optimal number theoretically and confirmed our predictions by the simulation results.

\section*{Results}

We used the coarse-grained model of PVA and PE \cite{meyer2002formation} to model a semicrystalline state. The studied systems contained chains of length up to $10^5$ monomer units. A distinctive feature of our work was the method of variation of entanglement density in initial conformations. We used relatively “natural” way of controlling the entanglement density by varying the reaction rate of polymerization with simultaneous crystallization, see the details in Methods and our previous work \cite{Petrov20}. The properties of the systems under study are summarized in Table\,\ref{table}. The $N_e50$ sample had an almost equilibrium entanglement density, while samples with larger $N_e$ values were in a nonequilibrium low-entangled state. We also studied a reference system $N_e53$*short that also had entanglement density close to equilibrium but consisted of much shorter chains. There is also a minor fraction of free solvent in all samples (around 10\%) as in the real polymer materials studied in experiments on fiber drawing.

\begin{figure}[htbp]
\centering
    \begin{subfigure}{0.49\textwidth}
	\includegraphics[width=\linewidth,height=\textheight,keepaspectratio]{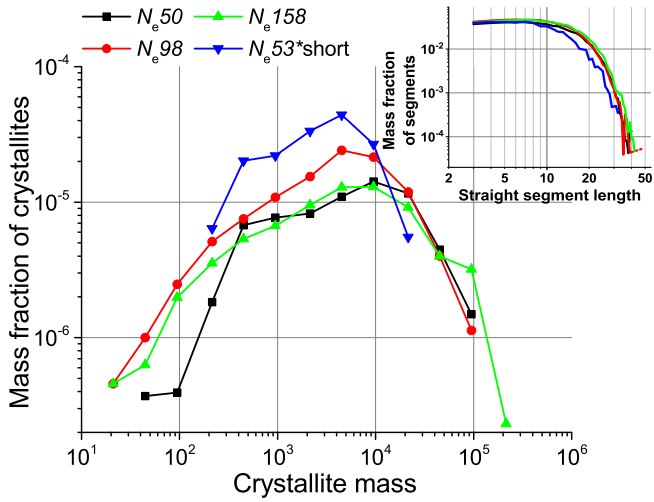}
	\caption{}
	\end{subfigure}
	\begin{subfigure}{0.49\textwidth}
	\includegraphics[width=\linewidth,height=\textheight,keepaspectratio]{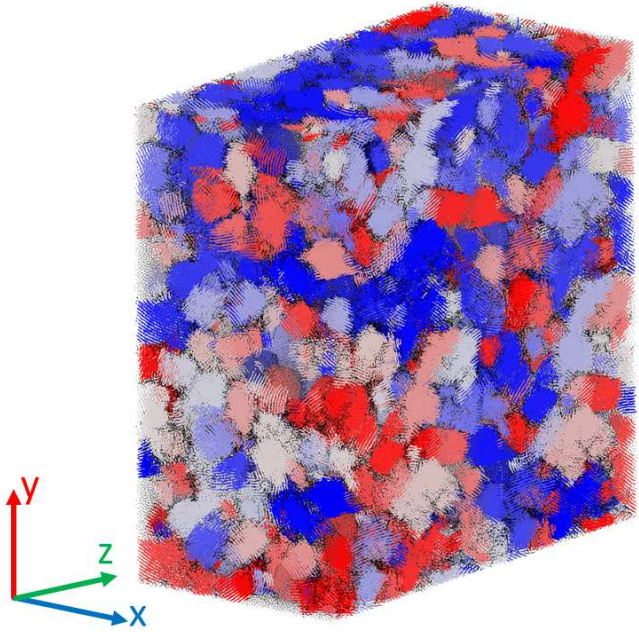}
	\caption{}
	\end{subfigure}
	\caption{(a) Crystallite mass fraction distributions for the undeformed systems. The inset demonstrates straight segment length distribution. (b) Snapshot of the simulation box before the deformation. Individual crystallites are shown in red--grey--blue palette in a random way. The sample is deformed along the x-axis.}
    \label{fig:crystmassdistr}
\end{figure}

Initially, we quantified the crystallite mass distributions in these systems (Fig.\,\ref{fig:crystmassdistr}a) and the length distributions of straight chain segments in crystallites (inset). Fig.\,\ref{fig:crystmassdistr}a suggests that entanglements influenced the crystallite mass distribution weakly; however, the overall crystallinity increased with entanglement length $N_e$ (Table \,\ref{table}). Fig. \ref{fig:crystmassdistr}b demonstrates the snapshot of a typical system before the deformation. The sample was deformed along the direction perpendicular to the larger plane (i.e. along the x-axis).

\begin{table}
\centering
\begin{tabular}{|l|p{1.5cm}|p{1.3cm}|l|l|l|}
\hline
&\thead{Acronym}&\thead{Average\\chain \\ length $N$}&\thead{Polymer \\ volume \\ fraction $f$}&\thead{Entanglement \\ length $N_e$}&\thead{Crystallinity}    \\
\hline
1      & $N_e50$ & 89995        & 0.900   & 50  & 0.59  \\
\hline
2      & $N_e98$ & 90026        & 0.900   & 98  & 0.66  \\
\hline
3      & $N_e158$ & 88539        & 0.885 & 158 & 0.69  \\
\hline
4      & $N_e53$*short & 902         & 0.902   & 53  & 0.51 \\
\hline
\end{tabular}
\caption{Properties of the systems under study.}
\label{table}
\end{table}

\subsection*{Stress--strain dependencies}

\begin{figure}[htbp]
\centering
    \begin{subfigure}{0.49\textwidth}
	\includegraphics[width=\linewidth,height=\textheight,keepaspectratio]{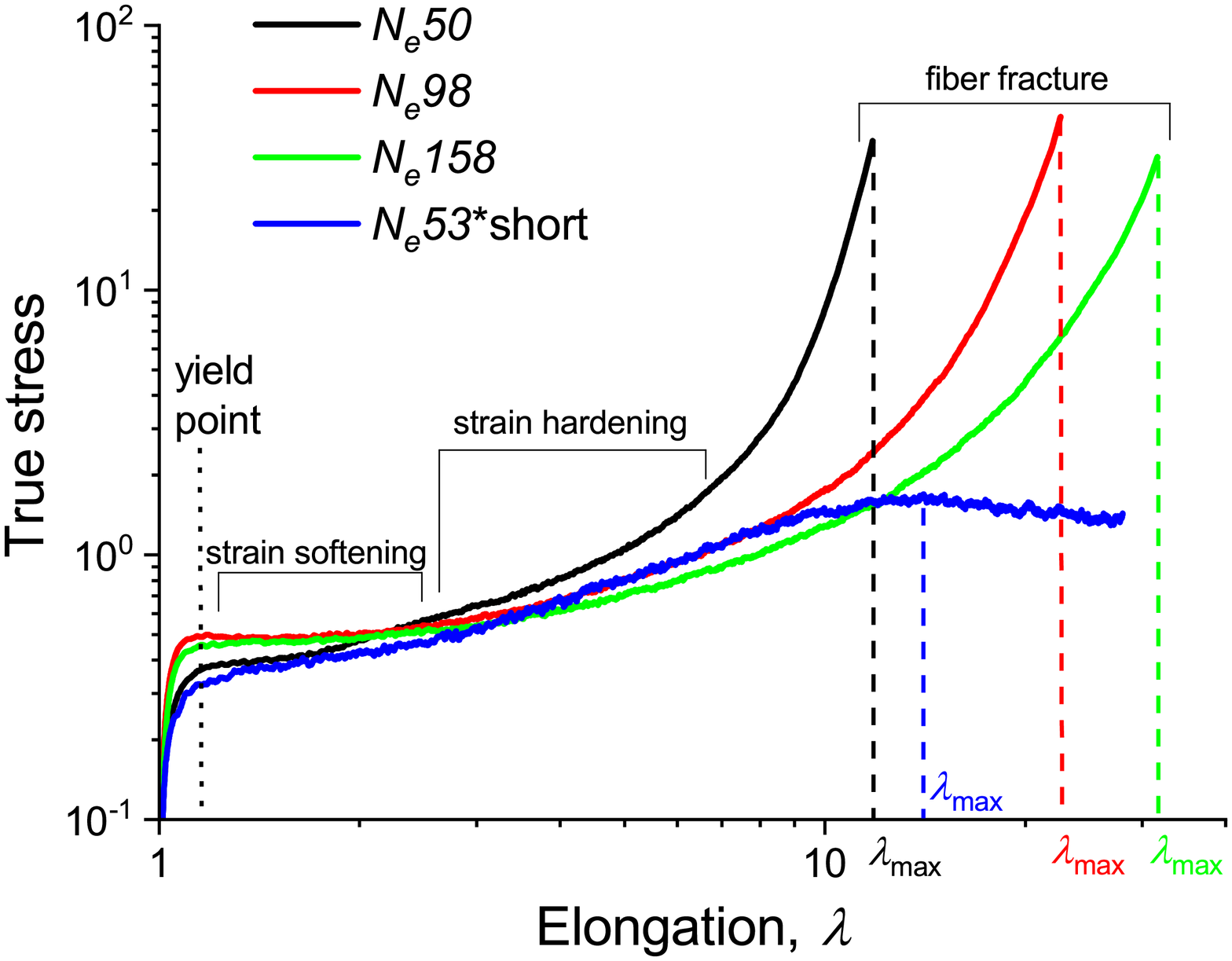}
	\caption{}
	\end{subfigure}
	\begin{subfigure}{0.49\textwidth}
	\includegraphics[width=\linewidth,height=\textheight,keepaspectratio]{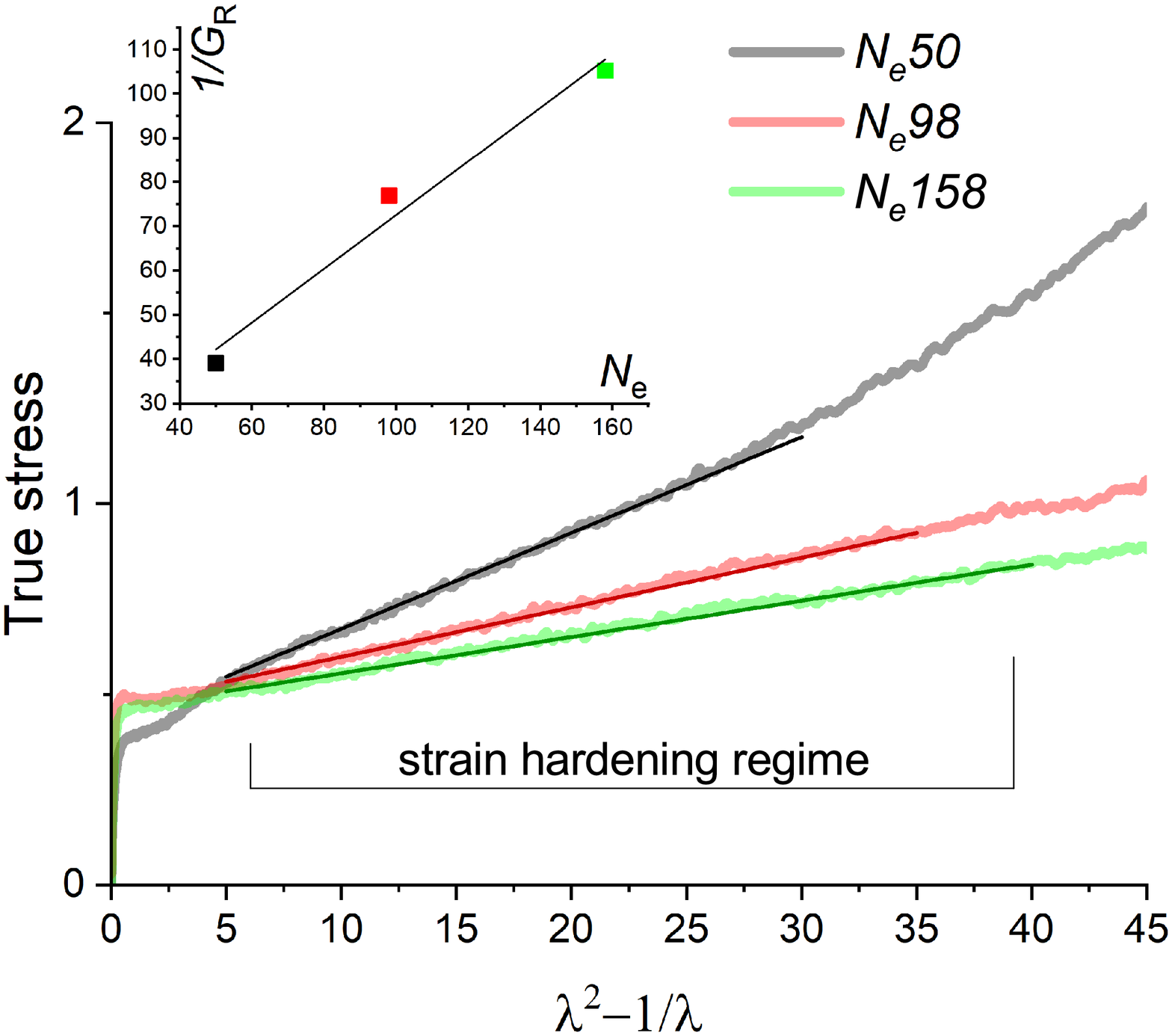}
	\caption{}
	\end{subfigure}
	\caption{a) True stress as a function $\lambda$. Dashed lines denote the moment of fracture except for the $N_e53^*short$ system that exhibits flow-like behavior for $\lambda>\lambda_{max}$. b) The dependency of true stress on $\lambda^2-1/\lambda$ for the systems with long chains. The inset shows the dependency of the inverse strain hardening modulus (slope of the black lines in the main plot) on the value of $N_e$ in the undeformed sample.}
    \label{fig:stressstrain}
\end{figure}

The simulations of uniaxial stretching were performed until the fracture that occurred at elongation ratios $\lambda$ up to 30; this value of $\lambda$ is approximately an order of magnitude larger than in the previous computer simulation studies. Here, $\lambda$ is defined as $\lambda(t)=L_x(t)/L_x(0)$, where $L_x(t)$ is the size of the simulation box along the $x$ axis at time $t$. We kept constant the strain rate, volume, and temperature. All systems were deformed along the x-axis with a selected strain rate $\Dot{\epsilon}=1.8\times10^{-5}\tau^{-1}$ that was large enough to perform simulations in a feasible time and small enough to make the stress-strain curves independent of the strain rate, see Methods and Supplementary Material for more details. Nose-Hoover thermostat preserved temperature at $T=0.5$, which is well below the crystallization point. True stress was calculated from the diagonal components of pressure tensor as $\langle p_{xx}\rangle -0.5 \langle p_{yy} + p_{zz}\rangle$. The resulting stress--strain curves are shown in Fig.\,\ref{fig:stressstrain}.

\begin{figure}[htbp]
\centering
  \includegraphics[width=\linewidth]{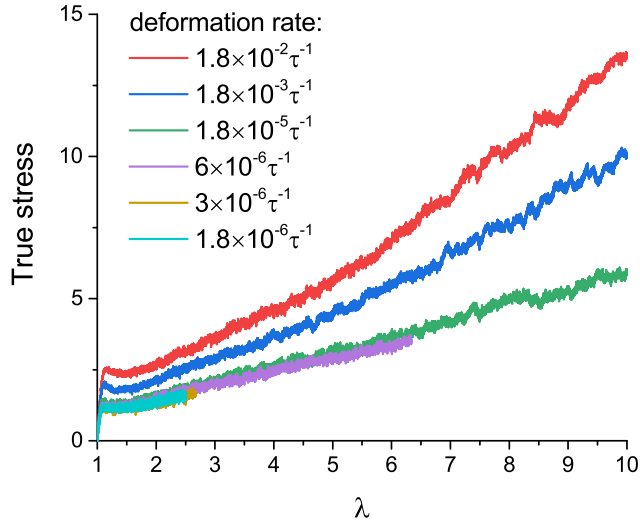}
  \caption{True stress-strain curves at different strain rates for the semicrystalline material obtained from cooling equilibrium CG polyethylene melt to the temperature $T=0.5$ (cooling rate was set to $10^{-6} \tau^{-1}$). The full procedure of preparation of this semicrystalline material is described in the Supporting Information to the work \cite{Petrov20}.}
  \label{fig:S2}
\end{figure}

At small values of $\lambda$, all curves exhibited a sharp increase (the Hookean elastic deformation) followed by the strain softening regime. We observed also a slight overshoot of the curve (yield point) that is strain rate-dependent (Fig. \ref{fig:S2}), so we will not focus on this point in this work. After the strain softening, we observed a gradual rise of the stress-strain curve indicating the strain hardening regime. In general, the presented stress response is very similar to the usual behavior of polymer materials under strain.

To assess the influence of entanglements on the stress-strain curves, we built the dependencies of true stress on $\lambda^2-1/\lambda$ (Fig. \ref{fig:stressstrain}b), which are the polymer network theory-driven coordinates. We observed that the stress-strain curves in those coordinates are well approximated by a linear dependency in the strain hardening regime, and its slope defines the so-called strain hardening modulus $G_{R}$. The $G_{R}$ values appear to be inversely proportional to the values of $N_e$ in an undeformed sample (see inset in Fig. \ref{fig:stressstrain}b), in full accordance with the affine network deformation model \cite{khokhlov1994statistical}. That observation demonstrated that even for semicrystalline polymers, the deformation in the strain hardening regime is controlled by entanglements acting similarly to cross-linkers in elastic polymer network. It is worth mentioning that the value of $G_R$ can be determined only at the beginning of strain hardening (Fig. \ref{fig:stressstrain}b). An increase of $G_R$ does not mean that the resulting fiber is stronger: true stress increases sharply with strain at the end of deformation (Fig. \ref{fig:stressstrain}a).

\begin{figure}
\centering
  \includegraphics[width=0.9\linewidth]{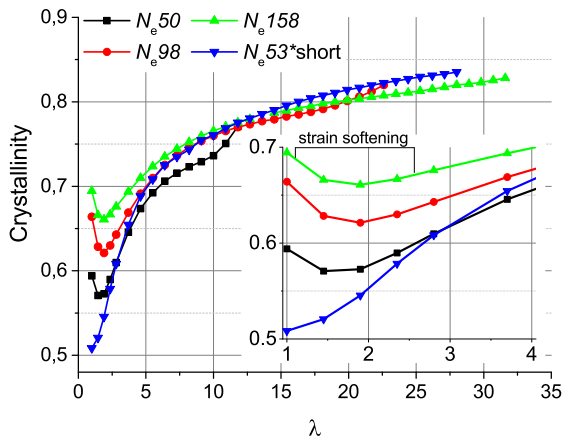}
  \caption{The dependency of crystallinity on $\lambda$.}
  \label{fig:crystallinity}
\end{figure}

The evolution of crystallinity during the deformation can explain the general trends of the stress response (Fig.\,\ref{fig:crystallinity}). For the systems with long chains, the dependencies exhibited the same pattern. First, crystallinity decreased at small $\lambda$. This clearly indicated the process of crystallite breaking at the beginning of deformation. Previous simulation studies \cite{lee2011plastic,jabbari2015plastic} supported this observation. Crystallinity reached a minimum at $\lambda\approx 1.5-1.9$ that roughly corresponded to the middle of the strain softening regime. After that, all systems exhibited strain hardening followed by recrystallization of broken crystallites and the formation of an oriented fiber; as a result, crystallinity gradually increased until the moment of fracture. The $N_e53^*short$ system, however, did not exhibit a decrease in crystallinity; in contrast, it monotonically increased with $\lambda$. This occurred probably due to smaller initial crystallinity in the deformed reference system, so crystallites could reorient without breaking.

\subsection*{Effect of entanglement length on the quality of the fiber}

For the systems with long chains, the strain hardening regime ended with a fracture. We observed a direct correlation between the maximum degree of deformation ($\lambda_{max}$) and the value of $N_e$ in the undeformed sample. The material with less entangled chains (with higher $N_e$) could be deformed to larger $\lambda_{max}$ (Fig. \ref{fig:stressstrain}a). However, for the $N_e53^*short$ system, the stress-strain curve started to slowly decrease after reaching some $\lambda_{max}$ value (Fig. \ref{fig:stressstrain}a). This behavior indicated the transition to the flow-like regime. Chains in the $N_e53^*short$ system contained an order of magnitude fewer entanglements per chain than the systems with long chains. Thus, a polymer material should have some critical number of entanglements per chain to be able to form a strong fiber and not to disaggregate during the process of drawing.

\begin{figure}[htbp]
\centering
    \begin{subfigure}{0.49\textwidth}
	\includegraphics[width=\linewidth,height=\textheight,keepaspectratio]{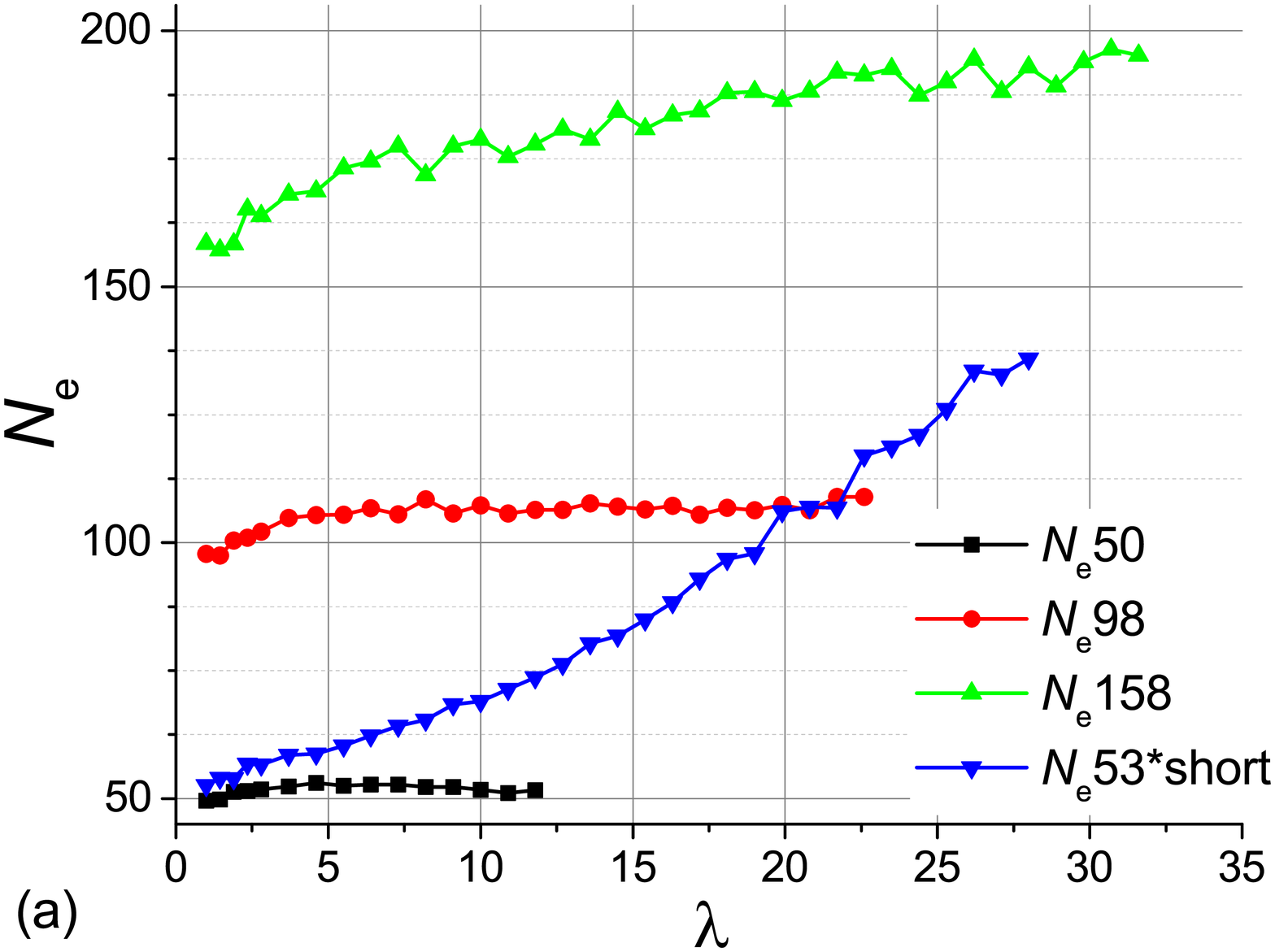}
	\end{subfigure}
	\begin{subfigure}{0.49\textwidth}
	\includegraphics[width=\linewidth,height=\textheight,keepaspectratio]{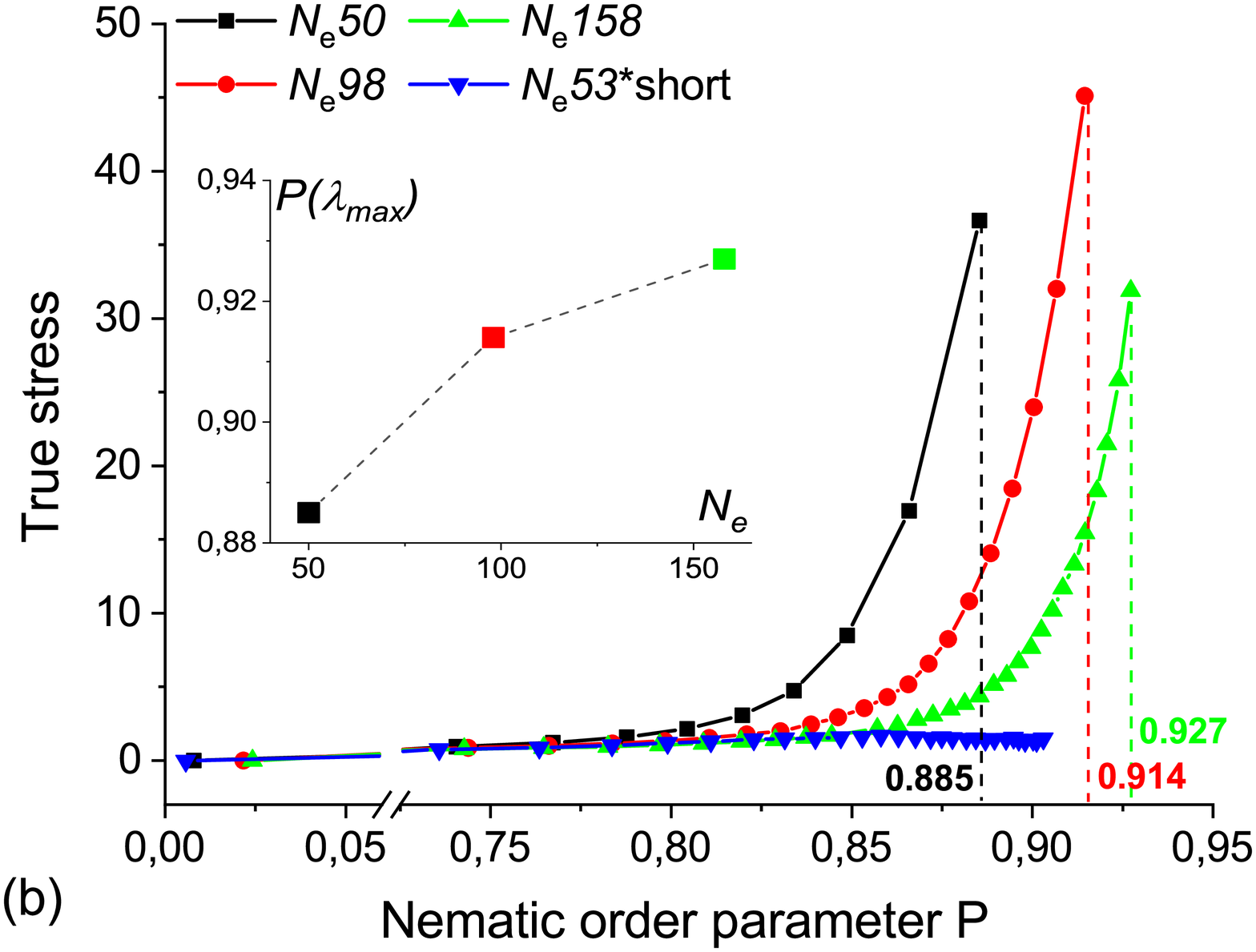}
	\end{subfigure}
	\caption{(a) The dependency of entanglement length on $\lambda$. (b) The dependency of true stress on the nematic order parameter $P$. Dashed lines represent the moment of fracture. Inset shows the dependency of $P$ in the fiber on the value of $N_e$ in the undeformed sample.}
	\label{fig:entanglements}
\end{figure}

The dependency of $N_e$ on $\lambda$ explains how $N_e$ can affect the fiber formation (Fig. \ref{fig:entanglements}a). We found that the short chains in the $N_e53^*short$ system greatly disentangled at the end of the deformation. In contrast, the $N_e(\lambda)$ dependencies for the systems with long chains exhibited plateau at large $\lambda$. Therefore, the key to maintain rigidity of the produced fiber is to stretch materials with chains that do not disentangle throughout the process of stretching.

However, polymers should be entangled not too strongly to form a highly oriented fiber. To quantify the degree of orientation of chain segments along the fiber axis, we calculated the nematic order parameter $P$, Fig. \ref{fig:entanglements}b.
The dependency of true stress on $P$ in the systems with long chains increased sharply at the end of deformation due to constraints imposed by entanglements prohibiting ideal orientation of chain segments. A very important observation from the Fig. \ref{fig:entanglements}b is that the value of $P$ in the fibers at the moment of fracture increased with $N_e$ (inset in Fig. \ref{fig:entanglements}b). Therefore, an increase of $N_e$ in the systems with long chains led to more oriented and, as a result, stronger fibers obtained after stretching the material.

To sum up, our data suggests the following conclusion: there is an optimal number of entanglements per polymer needed to maximize the fiber quality. When there are too few entanglements per chain in the material, polymers disentangle and do not form strong fibers after stretching (Fig. \ref{fig:entanglements}a). However, too strong entanglement leads to a decrease of the degree of chain orientation in the fiber (Fig. \ref{fig:entanglements}b). In the next section, we constructed an analytical theory describing how $N_e$ affects fiber formation and disentanglement and derived the aforementioned optimal number of entanglements per chain.

\subsection*{Analytical model}

In this section, we developed a simple theoretical model predicting the $\lambda_{max}(N_e)$ dependency that allowed us to predict the number of entanglements per chain in the fiber with the highest possible degree of chain orientation. First, we described the $\lambda_{max}(N_e)$ dependency for the materials able to form strong fibers. Second, we described the polymer materials that exhibit flow-like behavior for $\lambda>\lambda_{max}$ due to stress-induced disentanglement of chains.

We assessed the well-known characteristic in polymer theory: the average spatial distance $R$ between monomer units separated by $s$ monomer units along the chain, $R(s)$. We remind the reader that the chains in an equilibrium melt exhibit $R\propto s^{1/2}$ (random walk statistics). A single rod-like segment in a crystallite has $R \propto s$. The chain that folds densely like a space-filling curve (fractal globule) has $R \propto s^{1/3}$ \cite{grosberg1993crumpled}.

\begin{figure}[htbp]
\centering
	\begin{subfigure}{0.2\textwidth}
	\includegraphics[width=\linewidth,height=\textheight,keepaspectratio]{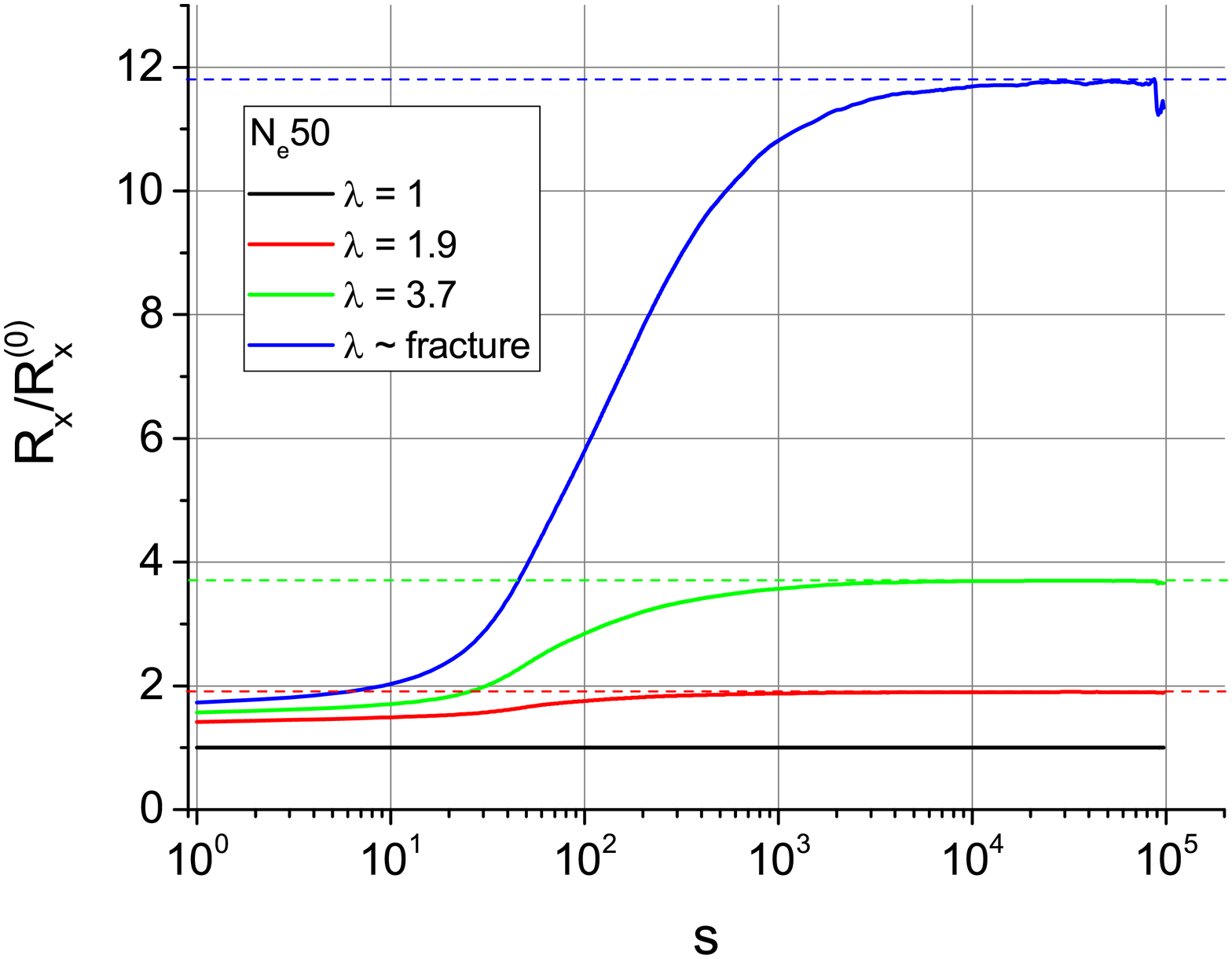}
	\subcaption{}
	\end{subfigure}
	\begin{subfigure}{0.2\textwidth}
	\includegraphics[width=\linewidth,height=\textheight,keepaspectratio]{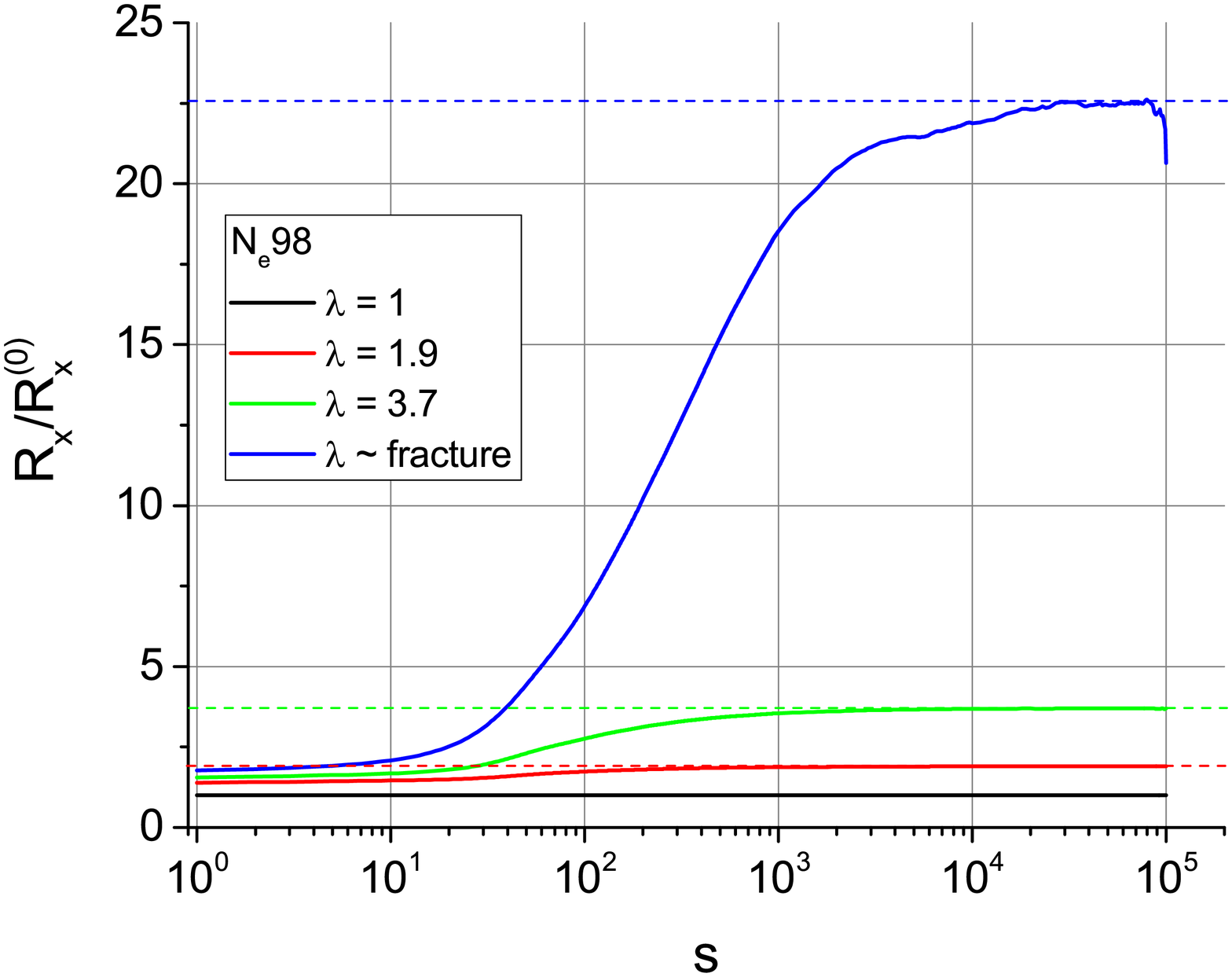}
	\subcaption{}
	\end{subfigure}
	\begin{subfigure}{0.2\textwidth}
	\includegraphics[width=\linewidth,height=\textheight,keepaspectratio]{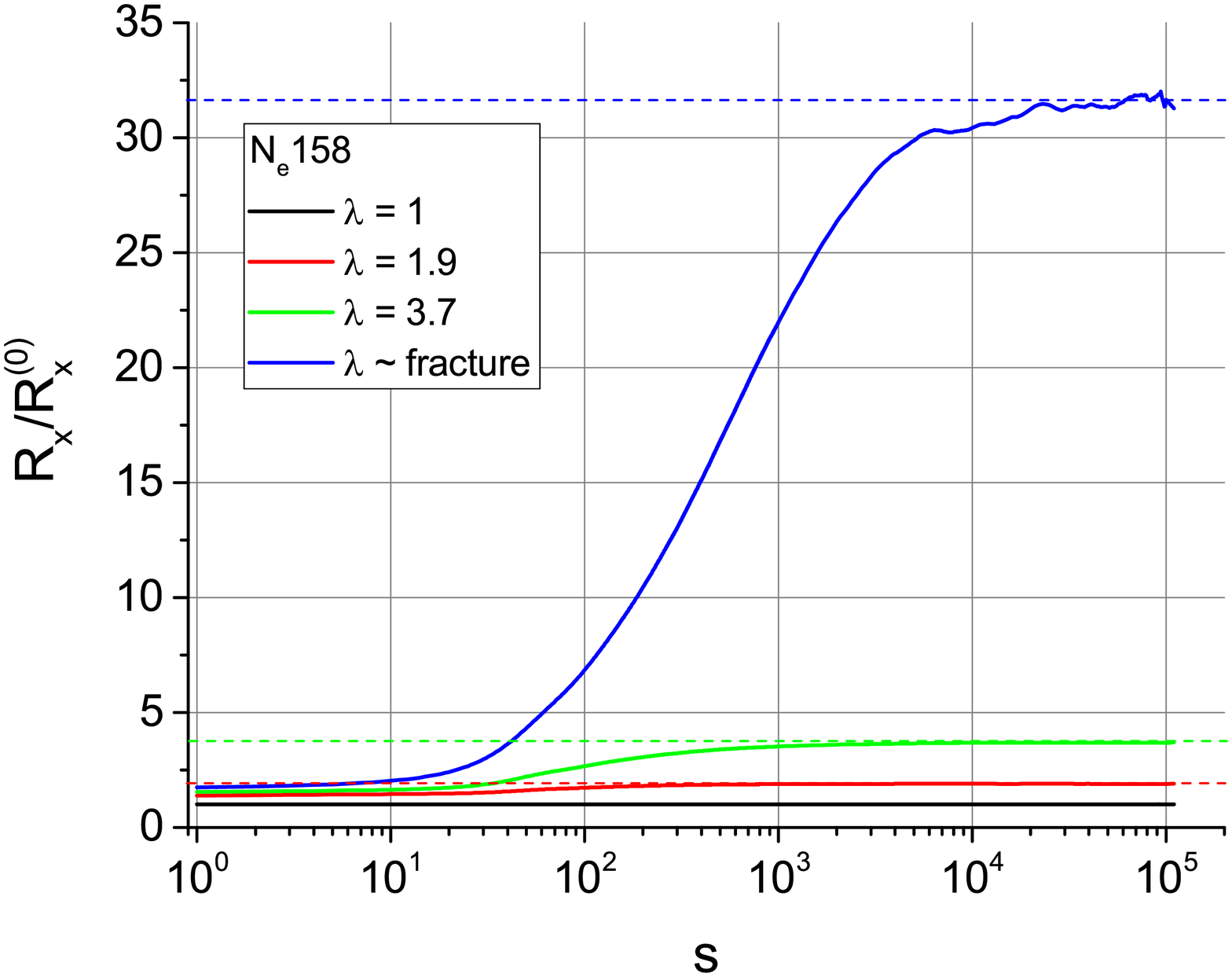}
	\subcaption{}
	\end{subfigure}
	\begin{subfigure}{0.2\textwidth}
	\includegraphics[width=\linewidth,height=\textheight,keepaspectratio]{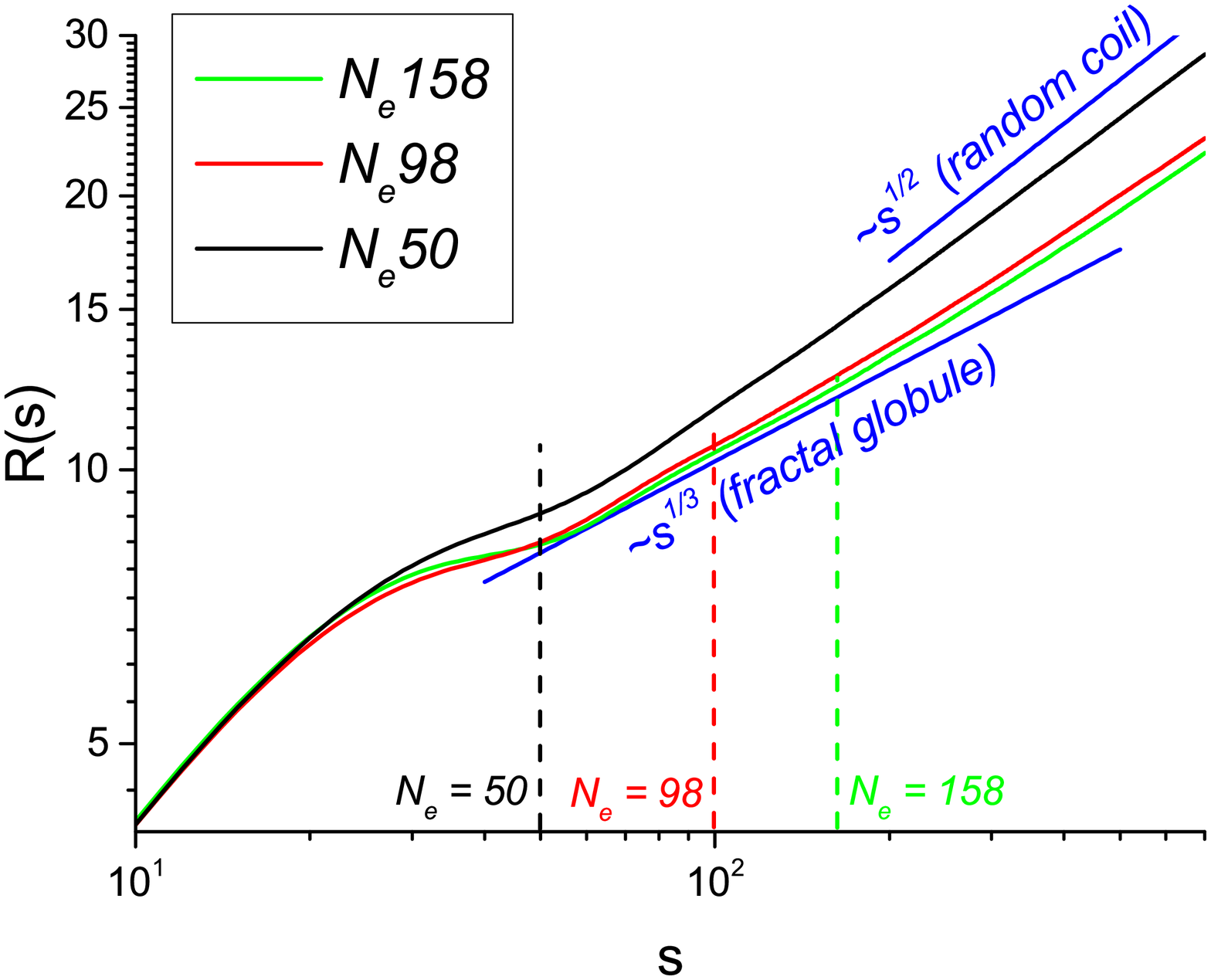}
	\subcaption{}
	\end{subfigure}
	\begin{subfigure}{0.2\textwidth}
	\includegraphics[width=\linewidth,height=\textheight,keepaspectratio]{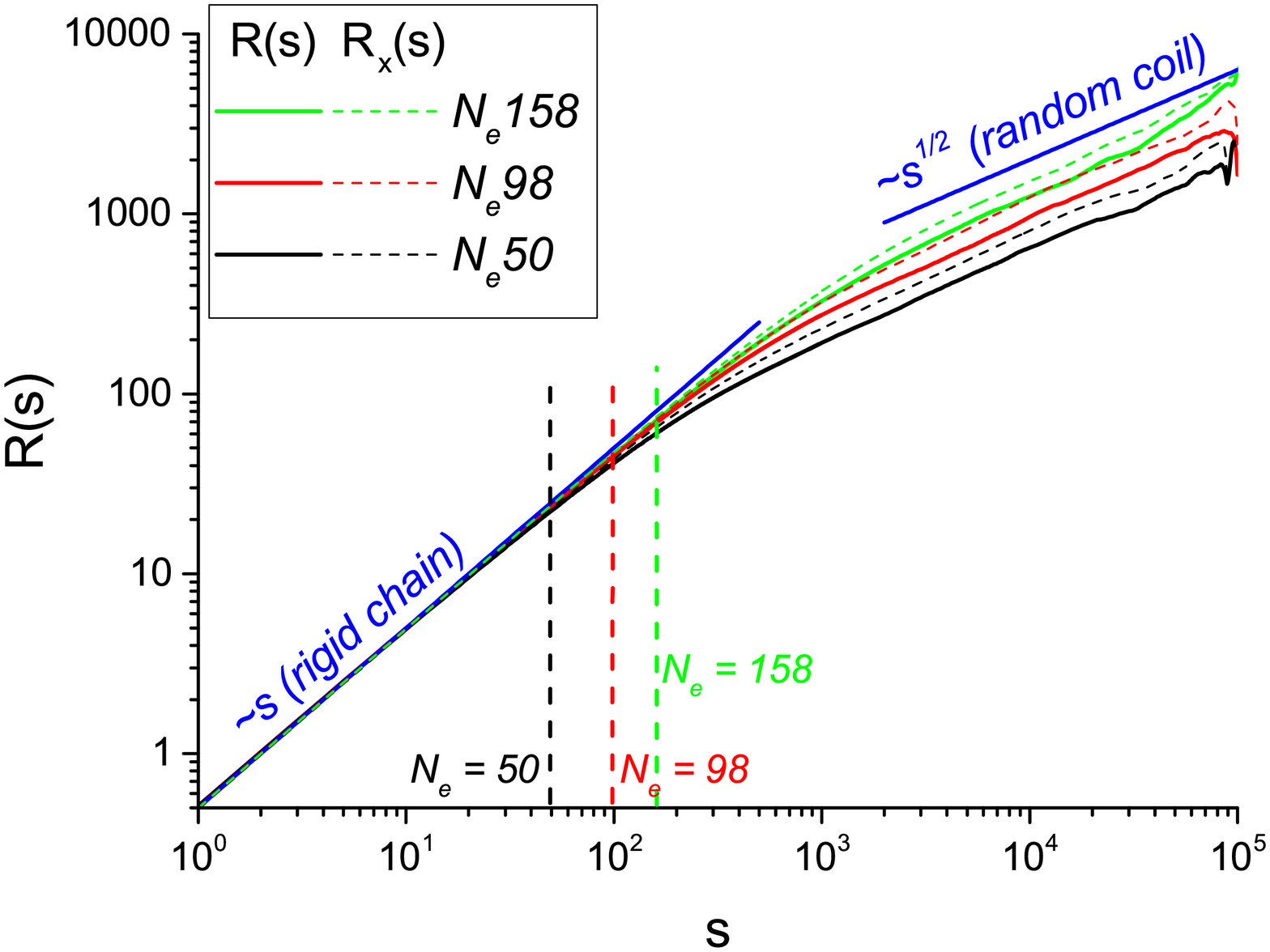}
	\subcaption{}
	\end{subfigure}
  \caption{The $R_{x}(s)$ dependencies at different $\lambda$ divided by the $R_{x}(s)$ dependency at $\lambda=1$ (denoted as $R_{x}^{(0)}(s)$) for the $N_e50$ (a), $N_e98$ (b), and $N_e158$ (c) systems. The dashed lines are $R_{x}(s)/R_{x}^{(0)}(s)=\lambda$ (affine deformation limit). d) The $R(s)$ dependencies for the systems with long chains in the non-deformed state. Blue lines are included as a guide for the eye. Vertical dashed lines represent the $s=N_e^{(0)}$ scale. e) The $R(s)$ and $R_x(s)$ dependencies for systems with long chains in the maximally deformed state.}
  \label{fig:S6}
\end{figure}

First, we consider the case when polymer chains do not disentangle during fiber formation. Thus, $N_e$ is constant due to extremely slow dynamics of long chains in a melt (the entanglement tube relaxation time scales as $\tau \approx N^{3.4}$, which is much larger than any accessible deformation time) hampered even more by the presence of the crystalline phase. In addition, we assume that the deformation is affine at the $s\propto N$ scale (Fig. \ref{fig:S6}a,b,c). As a result, $\lambda_{max} = R_{x}(s=N)|_{\lambda=\lambda_{max}}/R_{x}(s=N)|_{\lambda=1}$, where $R_x$ is the projection of the vector $R$ on the x-axis.

Let us describe the conformation of a polymer chain in an undeformed sample. A long polymer chain goes in and out of crystallites, resembling a random walk on a large scale and a densely folded "blob" on a small scale.Since monomer units are densely packed in a crystallite, the linear size of a "blob" scales as $R_b\propto N_{b}^{1/3}$, where $N_b$ is the average number of monomer units in a blob. Hence, taking into account that $R_{x}|_{\lambda=1} \propto R$, we obtain $R_x(s=N) \propto R(s=N)\propto N_{b}^{1/3} (N/N_{b})^{1/2}=N_{b}^{-1/6} N^{1/2}$ in an undeformed sample. Furthermore, formation of entanglements inside a crystallite is an unlikely event, since the chain conformation inside the "blob" is similar to the knot-free state of fractal globule \cite{grosberg1993crumpled}. Thus, we can expect $N_{b}\approx N_e$.  As a result, we obtain the following scaling relationship: $R_{x}(s=N)|_{\lambda=1}\propto N_e^{-1/6} N^{1/2}$.

Next, we estimate the dependence of $R_{x}(s=N)|_{\lambda=\lambda_{max}}$ value on $N_e$. We assume that the chain segments that formed a crumpled "blob" in a non-deformed sample tend to adopt a string-like conformation oriented along the x-axis. As a result, the $R(s)$ dependency behaves as $R(s)\propto s$ for $s<N_e$. On the other hand, the chain should still exhibit Gaussian behavior on the scale of the whole chain, yielding $R(s)\propto s^{1/2}$ for $s>>N_e$. Thus, a simple model of a single chain conformation in an oriented polymer fiber would be a random walk of straight segments of length $N_e$. The end-to-end distance of such a chain scales as $R(s=N)\propto N_e(N/N_e)^{1/2}=N_e^{1/2}N^{1/2}$. Since $R(s)\approx R_x(s)$ due to orientation of chains along the x-axis during deformation (Fig. \ref{fig:S6}e), we can write $R_{x}(s=N)|_{\lambda=\lambda_{max}}\propto N_e^{1/2}N^{1/2}$.

As a result, dividing the expressions derived in the previous paragraphs, we have $\lambda_{max}\propto N_e^{2/3}$. We can make another natural assumption to determine the coefficient of proportionality in the aforementioned expression. As $N_e$ tends to its minimum value of $N_e=1$, the material should be so rigid that it could not deform without undergoing fracture. Therefore, $\lambda_{max}(N_e=1)=1$. As a result, we arrive at the following equation relating $\lambda_{max}$ to $N_e$ measured in an undeformed semicrystalline material able to form a strong fiber:

\begin{equation}
\label{eq:1}
    \lambda_{max} = N_e^{2/3}
\end{equation}

It is worth mentioning that the expression similar to the Eq. \ref{eq:1} was derived previously for the behavior of rubber elastic materials and glassy polymers at large strains \cite{haward1968use,arruda1993three,myasnikova2011reactor}. The largest possible strain was predicted to be $\lambda_{max} = n^{1/2}$, where $n$ is the number of chain segments between two consecutive chemical crosslinks or points of entanglement. Therefore, the Eq. \ref{eq:1} is a modified version of the aforementioned expression in case of deformation of semicrystalline materials.

\begin{figure}
\centering
  \includegraphics[width=0.9\linewidth]{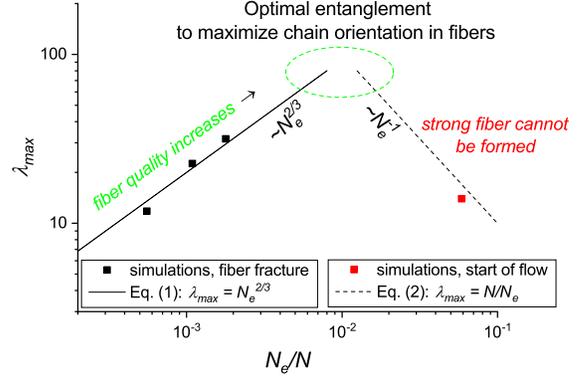}
  \caption{$\lambda_{max}$ as a function of $N_e/N$ ($N_e$ is measured in the undeformed samples). The black line and dots correspond to fibers, $\lambda_{max}$ is the strain at the moment of fracture. The red dot and the dashed line represent the systems that do not form strong fibers, $\lambda_{max}$ is the strain at which the material starts to exhibit flow-like behavior.}
  \label{fig:ne2}
\end{figure}

The second scaling model describes materials that exhibit flow-like behavior after reaching some critical stress and do not form strong fibers eventually. Our $N_e53^*short$ system consisting of short chains ($N\approx 10^3$) is an example of such a material. For this type of polymers, we define $\lambda_{max}$ as the strain at which the strain hardening ends, and the stress-strain curve reaches its global maximum (Fig. \ref{fig:stressstrain}a). As we mentioned before, materials start to flow at the end of strain hardening due to disentanglement of chains. Thus, $\lambda_{max}$ should scale as $\lambda_{max}\propto\lambda_C\times N/N_e$, where $N_e$ is the entanglement length in an undeformed sample. $N/N_e$ is the initial number of entanglements per chain, and $\lambda_C$ is the strain that causes removal of one entanglement per chain. Here we assumed that $\lambda_C$ is constant throughout the deformation. $\lambda_{max}$ should tend to unity in the $N/N_e\to 1$ limit, since chains would be in the disentangled state before the deformation in this limit. Hence, since $\lambda_C$ is a local characteristic that should be independent of $N$, it is also independent of $N_e$. Therefore, we derive the following relation in the regime when chains disentangle and cannot form a strong fiber: 

\begin{equation}
\label{eq:2}
    \lambda_{max} = N/N_e
\end{equation} 

We built the dependencies predicted by Eq. \ref{eq:1} and \ref{eq:2} and compared them with the simulation data (Fig. \ref{fig:ne2}). We observed a very good qualitative and even quantitative agreement despite the absence of any phenomenological adjustable parameters in our theory. To summarize, we developed a simple scaling theory demonstrating that there is an optimal number of entanglements per chain that maximizes $\lambda_{max}$ and, as a result, the degree of chain orientation along the fiber axis. We estimated this value to be around $N/N_e \approx 10^2$. 

\subsection*{Microscopic mechanism of deformation}

In this section, we analyzed the microscopic mechanisms of deformation, confirming the hypotheses stated in the previous section.

\begin{figure*}
\centering
\includegraphics[width=0.75\linewidth]{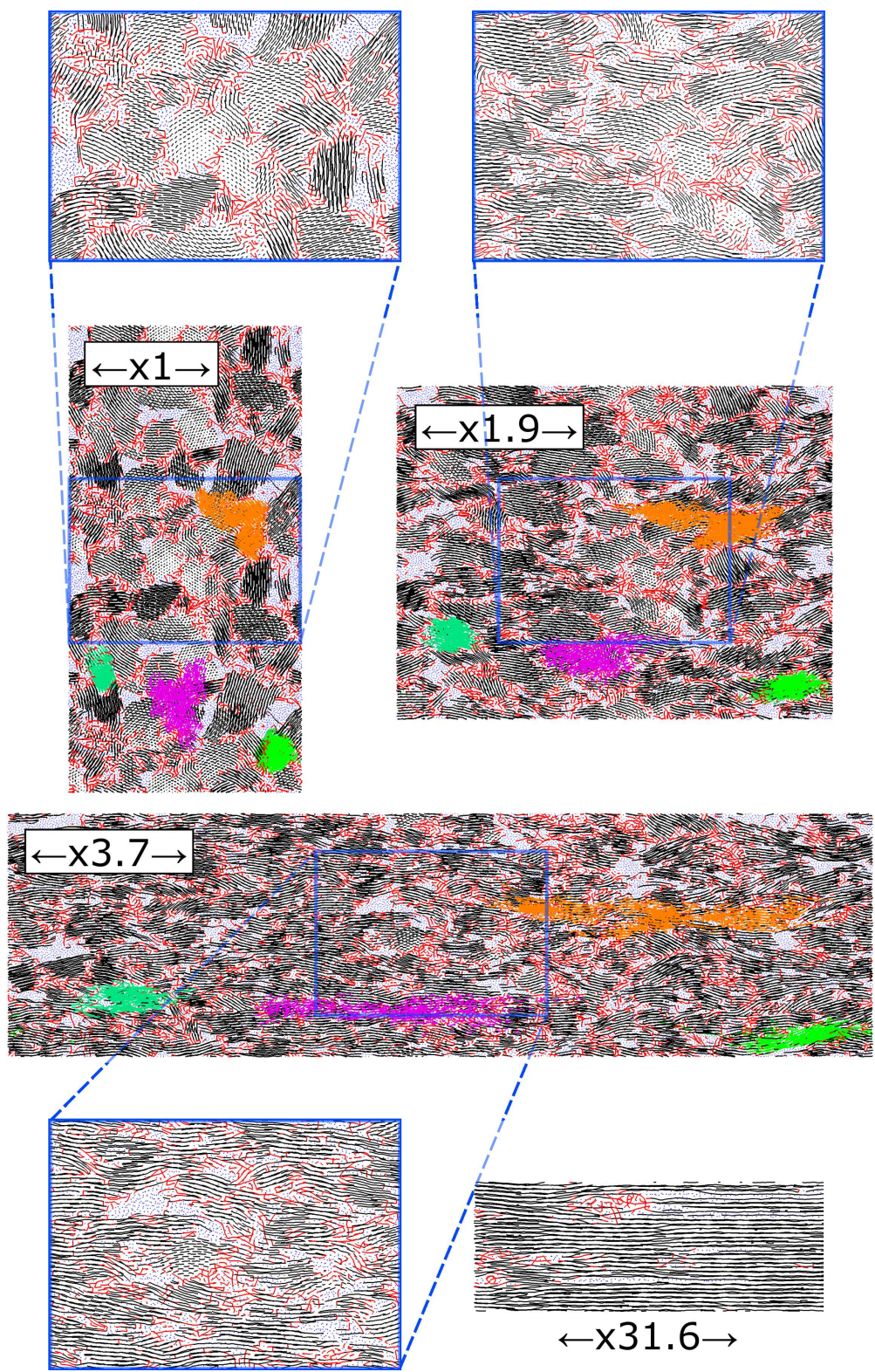}
  \caption{Full and detailed views of the system $N_e158$ at $\lambda=1$, $\lambda=1.9$, $\lambda=3.7$, $\lambda=31.6$ (detailed view only). Black and red lines represent crystallites and amorphous phase, respectively. Blue points represent solvent. In full view, beads forming four particular clusters in the non-deformed sample are colored in green, cyan, magenta, and orange. The mutual alignment of these objects remained unchanged during drawing, but each particular cluster changed its configuration and shape a lot. In detailed view, fragments of the same size are shown.}
  \label{fig:snapshotsall}
\end{figure*}

The detailed look at the snapshots of the systems with long chains helps to understand the underlying structural changes during the deformation. Fig.\,\ref{fig:snapshotsall} shows the typical crosscuts of the full systems before and during the deformation. This particular sample is the low-entangled one ($N_e=158$), but the other samples had a very similar structure. In addition, Fig.\,\ref{fig:snapshotsall} contains a detailed view of the evolution of crystalline and amorphous phases. At the beginning of deformation, crystallites change their orientation and break partially. After that, the process of recrystallization into the new elongated fibrils starts. Finally, the polymer chains form an oriented fiber that consists of the continuous crystalline phase and the amorphous inclusions containing entanglements. There is also a minor fraction of elongated solvent-rich voids located within the crystalline part of the oriented fiber.

\begin{figure}[htbp]
\centering
    \begin{subfigure}{0.45\textwidth}
	\includegraphics[width=\linewidth,height=\textheight,keepaspectratio]{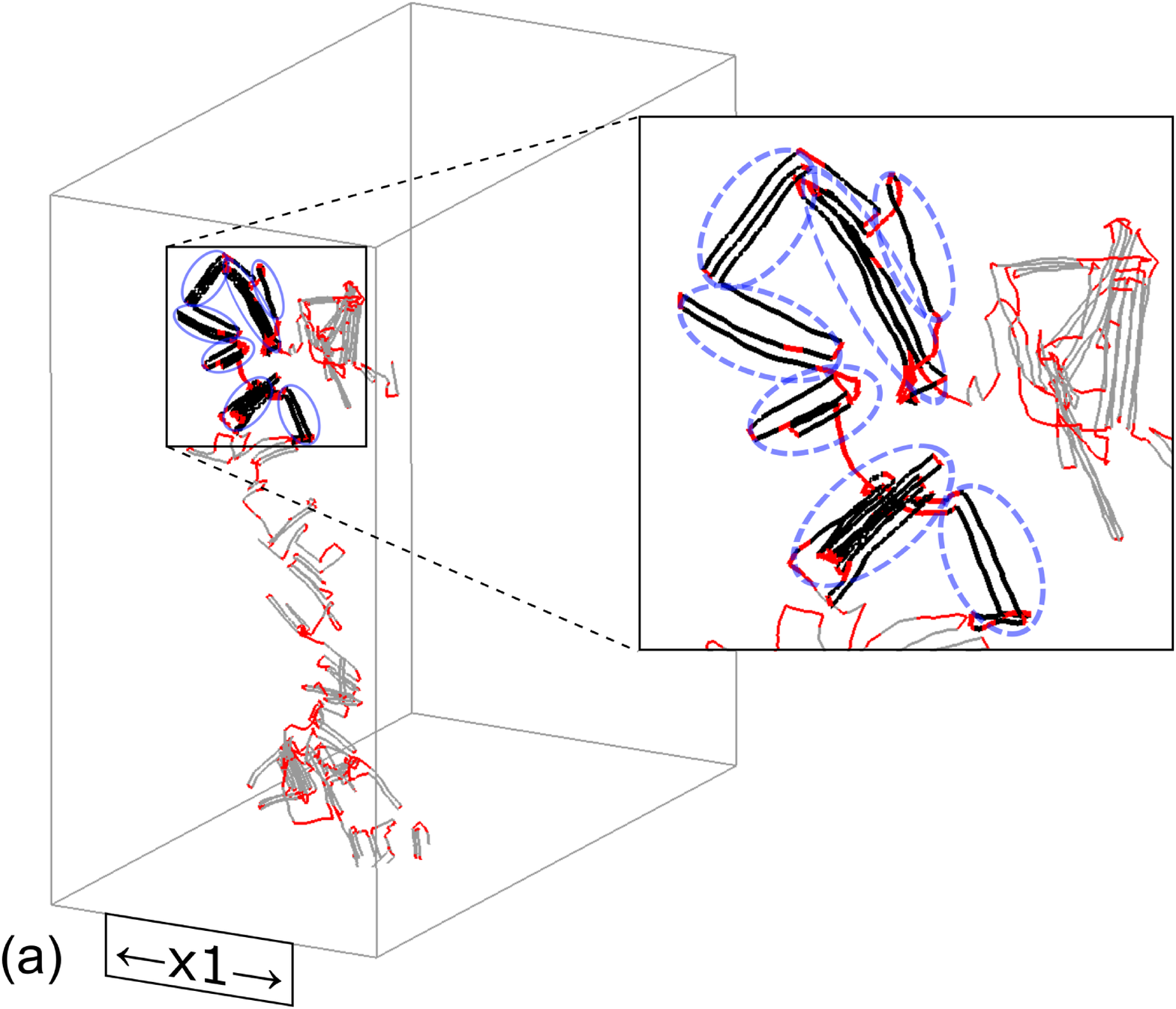}
	\end{subfigure}
    \begin{subfigure}{0.45\textwidth}
	\includegraphics[width=\linewidth,height=\textheight,keepaspectratio]{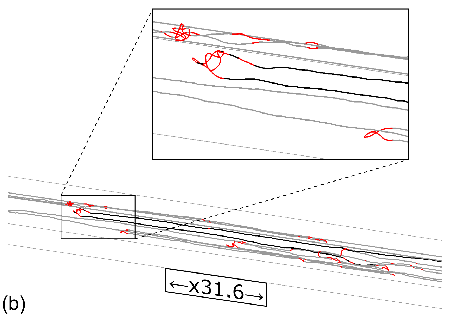}
	\end{subfigure}
    \begin{subfigure}{0.45\textwidth}
	\includegraphics[width=\linewidth,height=\textheight,keepaspectratio]{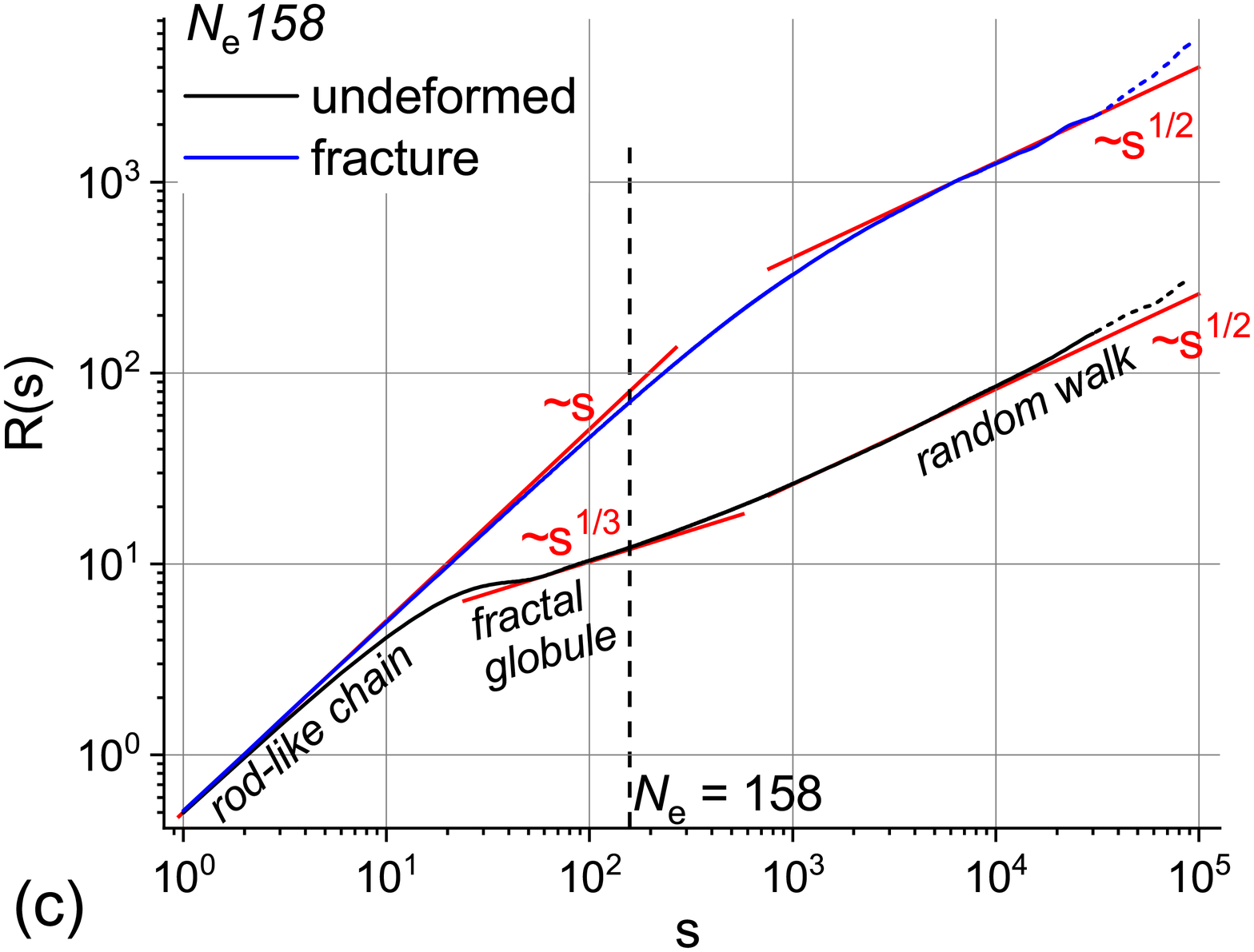}
	\end{subfigure}
	\caption{Snapshots of the $N_e158$ system at $\lambda=1$ (a) and $\lambda=31.6$ (b). Fragments consisting of 5000 particles in a single chain are shown in grey (crystalline parts) and red (amorphous parts). The 1000-beads segment are colored in black. At $\lambda=1$, crumpled blobs are circled in blue. (c) The $R(s)$ dependencies before the deformation and at the moment of fracture. Both curves demonstrate random walk statistics for $s>N_e$ and different folding at smaller scales.}
    \label{fig:rs}
\end{figure}

After that, we studied the mechanism of deformation on the level of individual chains. Fig. \ref{fig:rs}a shows the conformation of an individual chain in the non-deformed sample. We see that the chain travelled back and forth several times in one crystallite and then switched to another one. Therefore, the chain conformation can be viewed as a random walk of dense "blobs" as assumed in our theory. On the other hand, Fig. \ref{fig:rs}b demonstrates that the chain conformation is different in the fully stretched state. It resembles a random walk of straight segments making "turns" in amorphous (red) regions, and, obviously, all entanglements are also concentrated there.

\begin{figure}[htbp]
\centering
  \includegraphics[width=\linewidth]{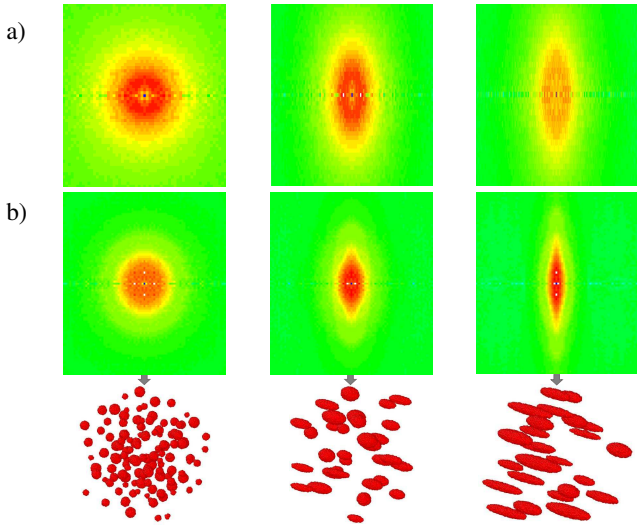}
  \caption{a) 2D SAXS pattern for the $N_e50$ system (taken from Figure 9 in the main text). Left - undeformed sample ($\lambda=1$), middle - $\lambda=1.9$, right - $\lambda=3.7$. b) 2D SAXS patterns calculated for the model systems shown below each plot. The number of particles in the model systems are approximately equal to each other (around $2\times 10^4$ in each structure). The left plot corresponds to the system of randomly positioned spheres with random radii (the average radius is 3.57). Middle and right plots correspond to the system of randomly positioned spheroids. Spheroids on the right plot are more elongated on average than the spheroids on the plot in the middle. The average ratio of the semi-major and semi-minor axes is 2.13 in the system in the middle and 5.24 in the system on the right. The semi-major and the semi-minor axes of spheroids are oriented along the x- and the z-axis, respectively.}
  \label{fig:S5}
\end{figure}

For numerical characterization of the single-chain conformations, we used the $R(s)$ dependency as in the previous section, Fig. \ref{fig:rs}c, \ref{fig:S5}d,e. Both the undeformed and the fully-stretched samples obey random walk statistics $R \propto s^{1/2}$ on a large scale, which is consistent with the analytical model and also with experimental SANS studies \cite{lopez2017molecular}. On a small scale, both samples exhibit the rod-like chain conformation ($R \propto s$), but up to different scales. For the oriented fiber, this scaling holds up to $s\approx N_e$. For the undeformed sample, however, the $R(s)$ dependency scales as $R \propto s$ only up to the value $s \approx 20$, which corresponds to the maximum of the straight segments length distribution (see inset in Fig. \ref{fig:crystmassdistr}a). The $R(s)$ dependencies for the undeformed samples follow the fractal globule scaling $R(s)\propto s^{1/3}$  on the intermediate scale (Fig. \ref{fig:rs}c, \ref{fig:S5}d). This observation supports the theoretical assumption of the dense entanglement-free folding of a single chain inside a crystallite and is consistent with the simulations of semiflexible chain collapse \cite{chertovich2014crumpled}.

\begin{figure}
\centering
  \includegraphics[width=1.0\linewidth]{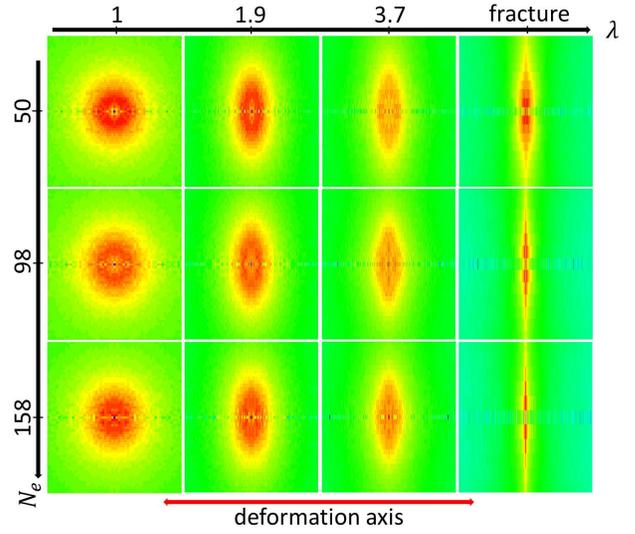}
  \caption{Calculated 2D SAXS patterns at various $\lambda$ for the systems with long chains.}
  \label{fig:saxs}
\end{figure}

The microscopic mechanism of deformation is often studied in experiments by employing the small-angle X-Ray scattering (2D SAXS) technique. In order to align our results regarding the mechanism of deformation with experimental data directly, we calculated the two-dimensional scattering patterns that roughly corresponded to 2D SAXS, Fig. \ref{fig:saxs}. In the non-deformed samples ($\lambda=1$), the patterns are isotropic and similar to the 2D SAXS data for the system of spheres with random positions and radii (Fig. \ref{fig:S5}). Thus, the regions of density contrast yielding the pattern in Fig. \ref{fig:saxs} roughly correspond to crystallites having random orientation and size before stretching.

\begin{figure}
\centering
  \includegraphics[width=0.9\linewidth]{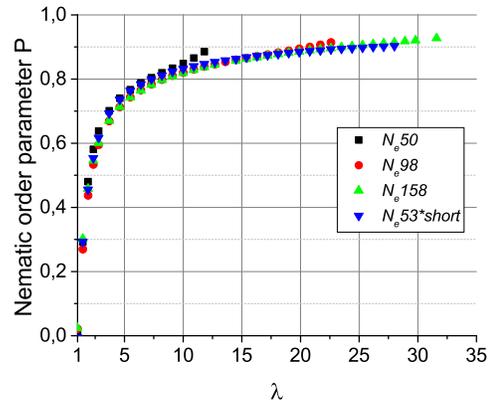}
  \caption{Nematic order parameter during drawing.}
  \label{fig:S4}
\end{figure}

During the deformation, the 2D SAXS patterns started to exhibit anisotropicity: the patterns contracted along the deformation axis and simultaneously elongated perpendicularly to the stretching direction (Fig. \ref{fig:saxs}). Experimental studies attributed this behavior to the formation of elongated oriented fibrils \cite{Macromol18}. 2D SAXS profiles for the test systems of randomly positioned spheroids confirmed this hypothesis: their elongation led to more anisotropic patterns elongated perpendicularly to the semi-major axes of the spheroids (Fig. \ref{fig:S4}).

\begin{figure}[htbp]
\centering
	\begin{subfigure}{0.46\textwidth}
	\includegraphics[width=\linewidth,height=\textheight,keepaspectratio]{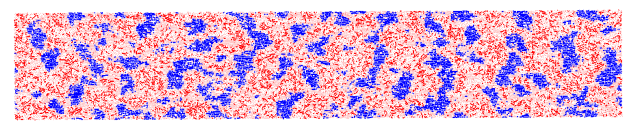}
	\subcaption{}
	\end{subfigure}
	\begin{subfigure}{0.46\textwidth}
	\includegraphics[width=\linewidth,height=\textheight,keepaspectratio]{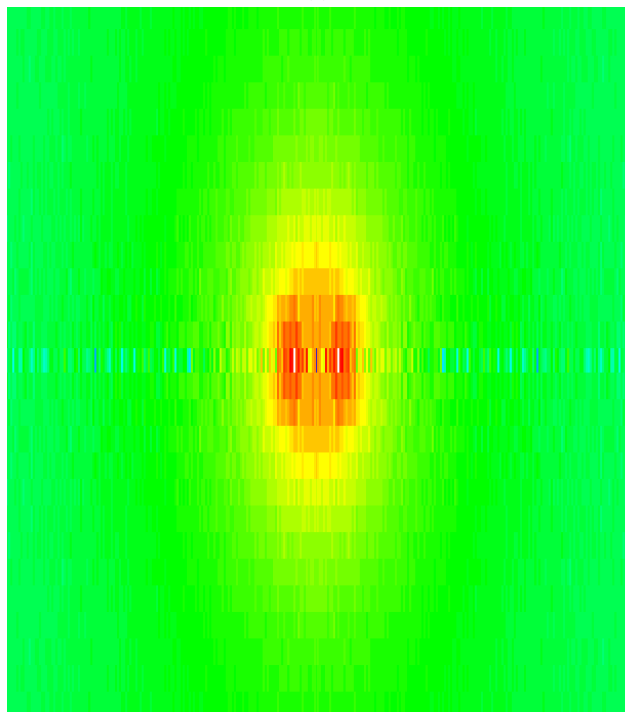}
	\subcaption{}
	\end{subfigure}
  \caption{Example of the hot-stretching process. (a) Snapshot of the system with long chains and equilibrium $N_e$ at $T=0.7$ deformed to $\lambda=5.05$. To prepare this sample, we carried out the fast polymerization and equilibration procedures described in the Section 1 and then cooled the system to $T=0.7$. After cooling, we performed deformation according to the procedure described in the main text. Crystalline and amorphous phases are colored blue and red, respectively. The deformation direction is horizontal. (b) The corresponding calculated 2D SAXS pattern. }
  \label{fig:S7}
\end{figure}

An important observation from the Fig. \ref{fig:saxs} is the absence of the shish-kebab structure that is often observed in experimental 2D SAXS data during hot-stretching of high molecular weight semicrystalline polymers \cite{tian2014lamellae,tian2015transition}. We observed only the streak-like pattern at the meridian of the 2D SAXS plot corresponding to formation of elongating fibrils ("shish"). There is no spot-like intensity maxima demonstrating the presence of crystalline domains oriented perpendicularly to the deformation axis ("kebab"). This phenomenon might occur due to the low temperature at which we carried out the deformation simulations: $T=0.5$ roughly corresponded to $2\text{C}$ \cite{Petrov20}, which is well below the usual temperatures used in tensile deformation experiments. We propose that, unlike the classical hot-stretching, the low-temperature solid-state deformation of high molecular weight polymers (the so-called "cold drawing") does not proceed via the "shish-kebab" structure formation. To confirm this hypothesis, we performed additional simulations of hot-stretching and observed a well pronounced "shish-kebab" structure and the characteristic 2D SAXS pattern with intensity maxima, see Fig. \ref{fig:S7}. We believe that all conclusions about entanglements established in this work in the case of cold drawing are also applicable to the process of hot-stretching; however, a detailed comparison of these two processes is beyond the scope of this paper.

\section*{Discussion and conclusions}

We carried out large-scale computer simulations of oriented polymer fiber formation. Long enough polymer chains did not disentangle during sample drawing, and an oriented fiber was formed after strain hardening. In this regime, a decrease of the entanglement density led to the formation of a more oriented and thus stronger fiber. At the same time, the samples having too few entanglements per chain started to exhibit flow-like behavior in the middle of strain hardening and, as a result, did form a strong fiber. We built a theory describing the processes of fiber formation and disentanglement. We discovered that there is an optimal ratio between the total polymer length $N$ and the entanglement length $N_e$ necessary to obtain fibers with the highest degree of chain orientation. Using our theory, we estimated the optimal $N/N_e$ ratio to be around $10^2$. On the other hand, experimental studies suggest \cite{christakopoulos2021melting,Rastogi11} that the equilibrium $N/N_e$ ratio for ultrahigh molecular weight polyethylene (UHMWPE) is around $3\times 10^3$ (for the polymers with $M_w\approx 5\times 10^6$ g/mol). Homogeneous polymerization with simultaneous crystallization was able to reduce the entanglement density approximately 3-fold in the synthesized nascent polymer having such molecular weight \cite{Rastogi11}. Therefore, the lowest $N/N_e$ ratio achieved during the solvent-free route of UHMWPE fiber production is approximately $10^3$, which is an order of magniture larger than the predicted optimal value. Thus, we believe that synthesis of UHMWPE having the same chain length ($M_w\approx 5\times 10^6$ g/mol) but an order of magnitude larger entanglement length will lead to production of the fiber of maximal strength.

A minor but important result of our work is the direct assessment of the strain hardening modulus $G_{R}$ that determines malleability of polymer materials during strain hardening. We found out that $G_{R}$ was inversely proportional to the value of $N_e$ in an undeformed sample (Fig. \ref{fig:stressstrain}b). This finding proves directly the well-known hypothesis about the key role of amorphous entanglement network during late stages of polymer deformation.

In addition, for the first time in computer simulation studies, we calculated more or less realistic 2D SAXS patterns to characterize the microscopic mechanism of fiber formation. Using the 2D SAXS data, we discovered the fundamental difference between the mechanisms of fiber production during cold drawing and hot-stretching. Unlike the hot-stretching, the deformation at low temperature proceeded not via the "shish-kebab" mechanism, but rather through formation of gradually elongating fibrils. 

Our results are applicable to the case of UHMWPE fiber formation, since we used the coarse-grained model of polyethylene, and the deformed samples contained ultra-long chains similarly to the real UHMWPE samples. At the same time, we believe that our findings are applicable to any partially crystallized polymer material consisting of linear chains due to a rather general nature of both the coarse-grained model and the theoretical approach.

\section*{Methods}

\subsection*{Preparation of semicrystalline polymer samples}

First, we performed CG MD simulations of polyethylene homogeneous polymerization in poor solvent below the crystallization temperature using the same model as in ref. \cite{Petrov20}. Polymerization reaction was modeled using the stochastic ``mesoscale chemistry'' model \cite{berezkin2011simulation,gavrilov2015thermal,rudyak2017complex} in NPT ensemble with pressure $P=8.0$ (analogous to 1 atm) and temperature $T=0.5$ (above the temperature of glass transition, but below the temperature of crystallization) up to conversion degree $f\approx 0.9$.
We prepared four different systems. Three of them contained $9\times 10^5$ particles, including 9 initiator particles. These systems were polymerized at various reaction rates, which directly affected the resulting entanglement length $N_e$. Thus we obtained three semicrystalline systems composed of chains of length $N\approx9\times10^4$ beads and $N_e$ from 50 (close to $N_e$ in equilibrium polyethylene melt \cite{Petrov20}, obtained by a procedure of fast polymerization, Supplementary Material, Section 1) to 158 (see Table\,\ref{table}). The fourth system was a reference one and consisted of relatively short chains. It contained $184\times10^3$ particles, including 184 initiators, leading to 184 chains of length $N\approx900$ beads and $N_e=53$. It is worth mentioning this system was much less entangled than the others: the number of entanglements per chain $N/N_e$ was $N/N_e\propto 10^1$ in it, while in the systems with long chains $N/N_e\propto 10^2-10^3$.

Finally, we replicated each system along $y$ and $z$ directions and made large systems with dimensions $L_x\times 2L_y \times 2L_z$, containing from $7.36\times10^5$ particles (reference system) to $3.6\times10^6$ particles (other systems). This replication allowed us to reach larger degrees of deformation and thus to investigate the behavior of the samples from the very beginning of deformation up to the moment of fracture. Fig.\,\ref{fig:crystmassdistr}b shows the principal structure of large systems (on example of the system with $N_e=158$). 

\subsection*{Tensile Deformation Procedure}
We simulated the process of volume-conserving tensile deformation by using the standard procedure implemented in LAMMPS software. Temperature, volume interaction parameters, and bond potentials were identical to those used during polymerization \cite{Petrov20}.

We chose the strain rate $\Dot{\epsilon} = (L_x-L_{x0})/(L_{x0}\Delta t)$ as follows. Our goal was to perform tensile deformation slowly enough to prevent formation of cavities and simultaneously to carry out simulations in feasible time. In order to determine the optimal strain rate, we performed tensile deformation of a semicrystalline material obtained by cooling of the equilibrium CG polyethylene melt to the temperature $T=0.5$ (Fig. S1). We determined that the stress-strain curves were almost identical when the strain rate was less than or equal to $1.8\times 10^{-5}\tau^{-1}$ (reverse time units). Similar strain rates were used in other works on deformation simulations \cite{jabbari2015plastic,jabbari2017role,hoy2007strain}. In addition, we believe that using a smaller strain rate would not significantly affect the results, since the high degree of crystallinity and the large chain length complicates rearrangement of the system and results in slow and saturating disentanglement dynamics. Therefore, we chose this value of strain rate and used it in all our simulations.

\begin{figure}[htbp]
    \centering
	\begin{subfigure}{0.2\textwidth}
	\includegraphics[width=\linewidth,height=\textheight,keepaspectratio]{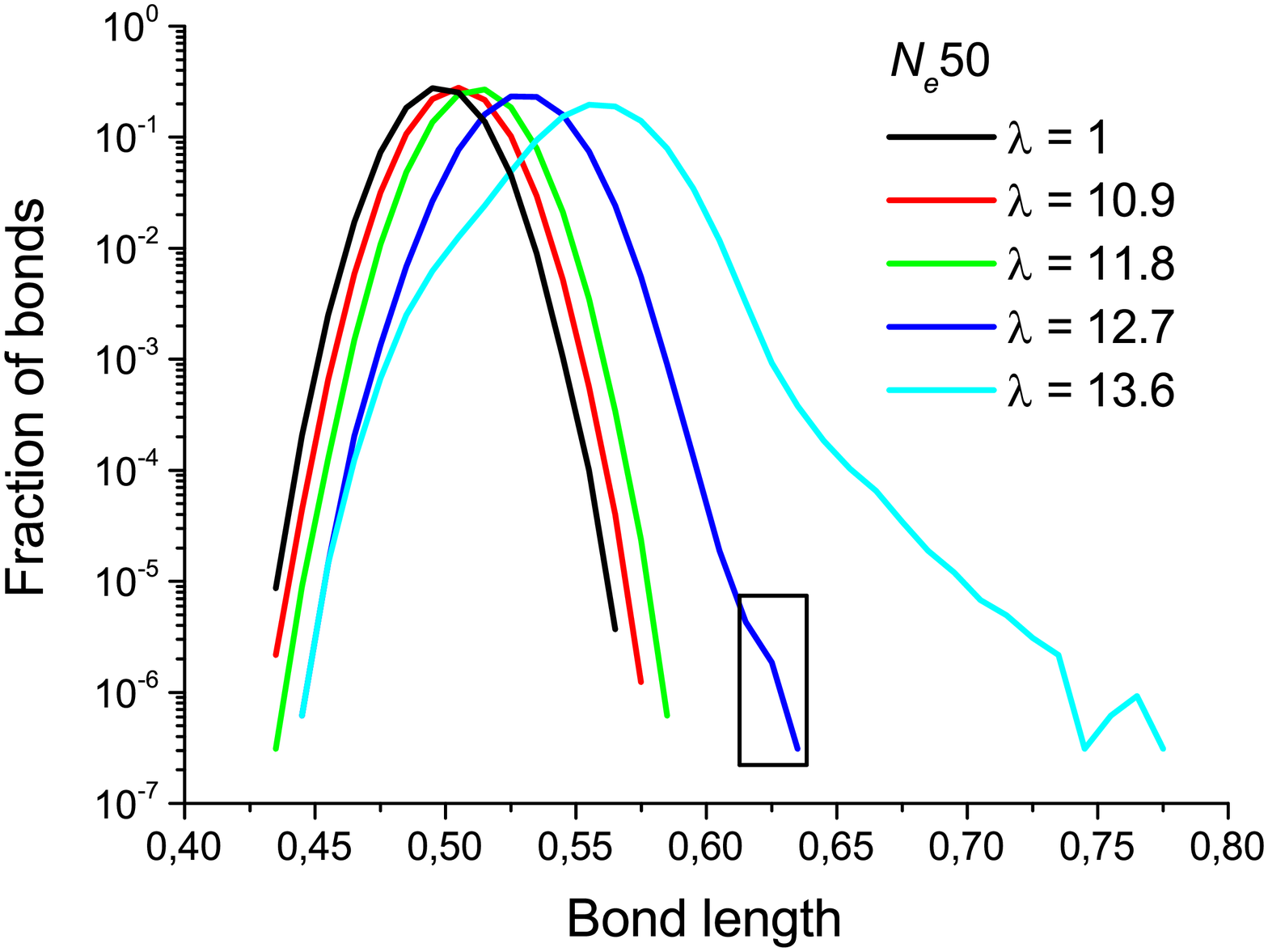}
	\subcaption{}
	\end{subfigure}
	\begin{subfigure}{0.2\textwidth}
	\includegraphics[width=\linewidth,height=\textheight,keepaspectratio]{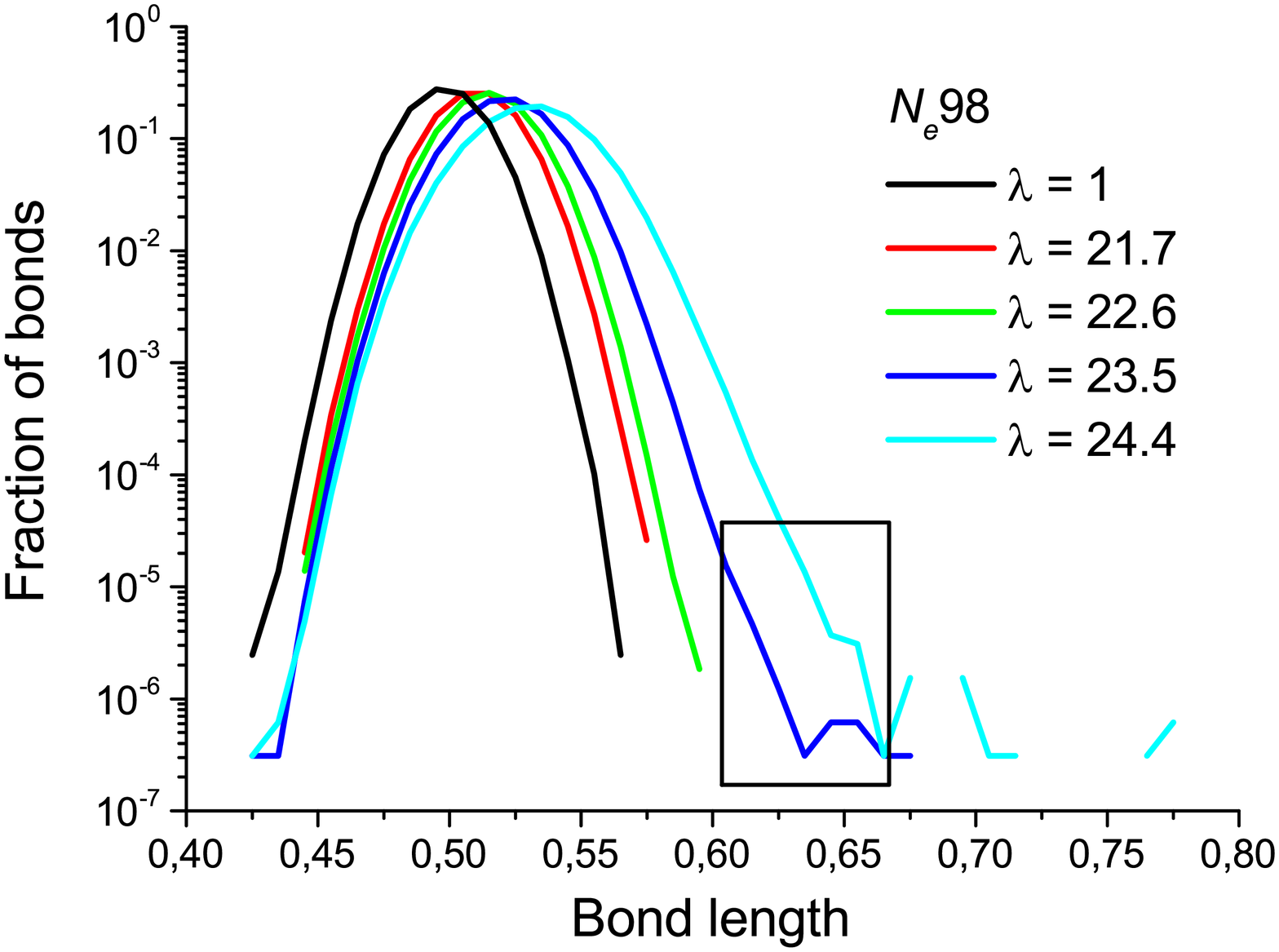}
	\subcaption{}
	\end{subfigure}
	\begin{subfigure}{0.2\textwidth}
	\includegraphics[width=\linewidth,height=\textheight,keepaspectratio]{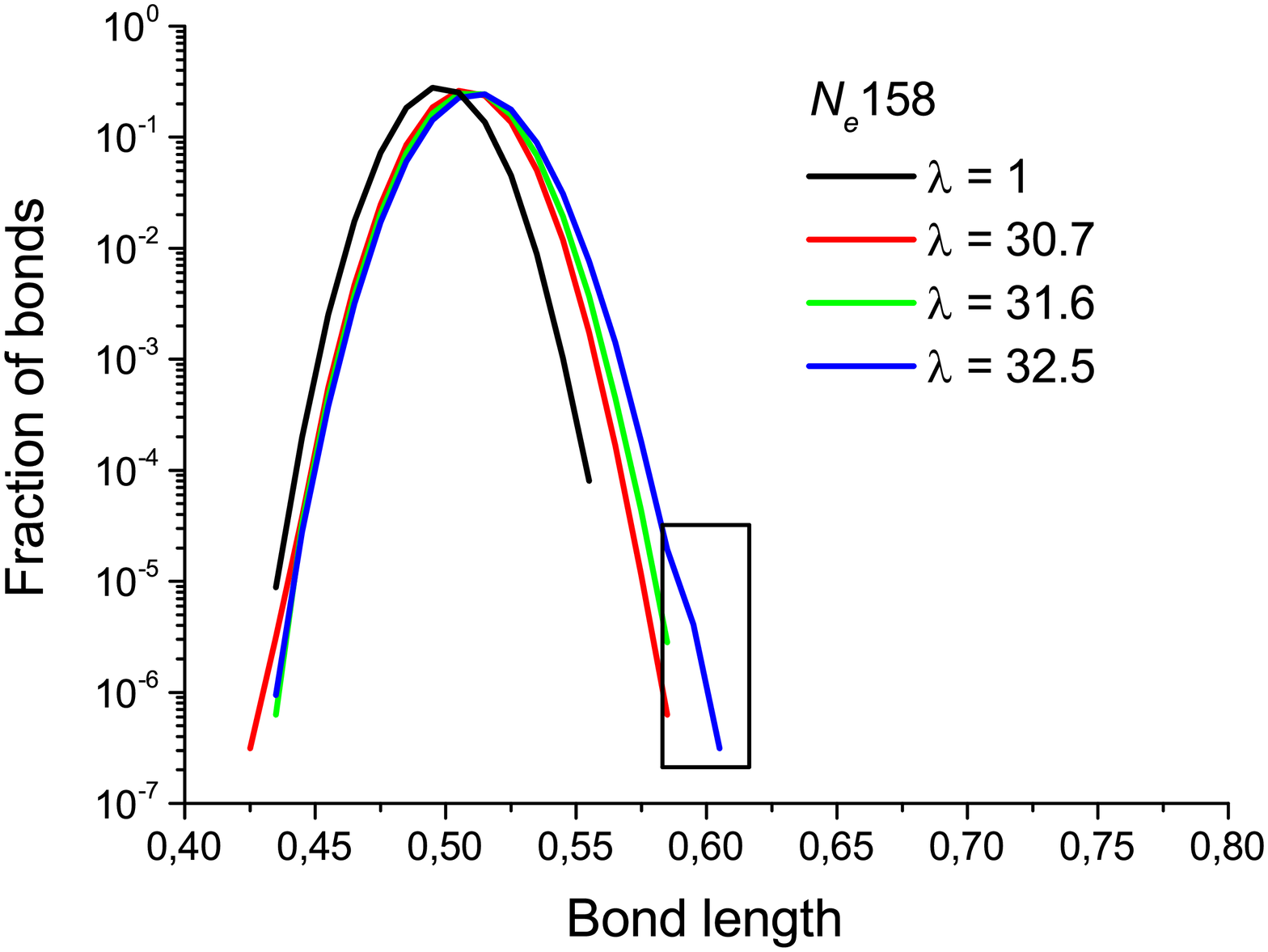}
	\subcaption{}
	\end{subfigure}
	\begin{subfigure}{0.2\textwidth}
	\includegraphics[width=\linewidth,height=\textheight,keepaspectratio]{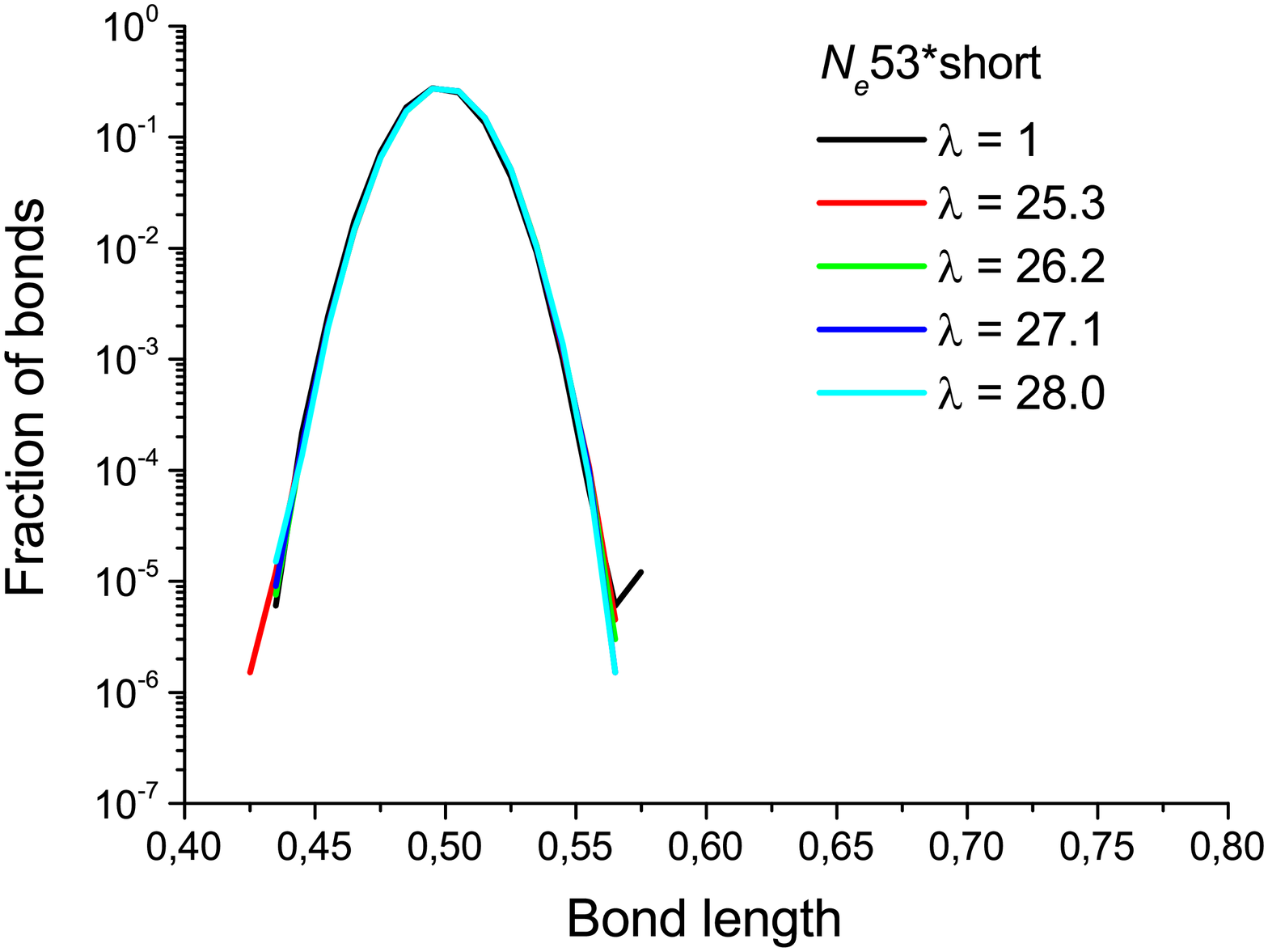}
	\subcaption{}
	\end{subfigure}
	\begin{subfigure}{0.2\textwidth}
	\includegraphics[width=\linewidth,height=\textheight,keepaspectratio]{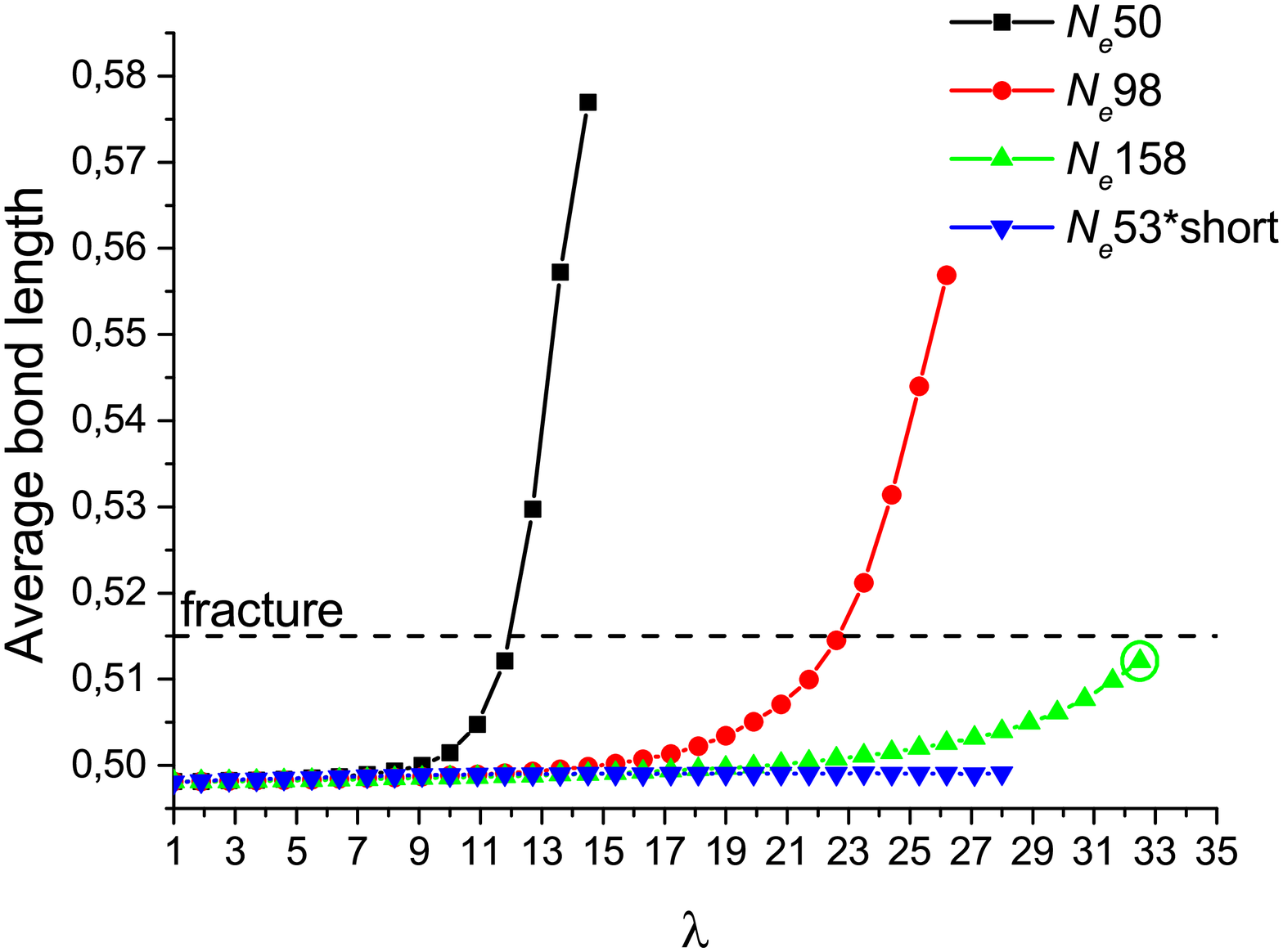}
	\subcaption{}
	\end{subfigure}
    \caption{a)-d) The bond length distributions in the systems. We treated the systems with heavy-tailed distributions as fractured. Heavy tails in the systems at the moment of fracture are enclosed within the black rectangles. e) The dependencies of the average bond length in the system on the degree of deformation $\lambda$ in different systems. The horizontal dashed line represents an approximate bond length threshold, beyond which the fracture occurred. The green circle represents the only exception: we treated this system as fractured based on the bond length distribution in this system (Fig. c).}
    \label{fig:S3}
\end{figure}

We performed simulations of deformation until fracture of the studied semicrystalline material. To determine the moment of fracture for a semicrystalline material, we analyzed distributions of bond lengths in the system (Fig. \ref{fig:S3}a-d). A heavy tail of those distributions indicated the appearance of a significant amount of anomalously long bonds. Therefore, we considered the fracture as a moment when the heavy tail appeared in the bond length distribution.

\subsection*{Calculation of Analyzed Quantities}

First, we calculated the degree of crystallinity (or simply "crystallinity"). This quantity characterized the average fraction of crystallized part in a semicrystalline material. For this, we evaluated for every bond the local fraction of co-directional neighbour bonds $\kappa$ in radius $r^\kappa_\text{cut}=2$. Two bonds were considered co-directional if they form angle less than 15$^\circ$. Particles forming bonds with $\kappa\geq0.3$ were considered as parts of crystallites. Thus the degree of crystallinity was calculated as ratio between number of such particles and total number of particles. 

Nematic order parameter was calculated as $P=1/2\left<(3\cos(\hat{\mathbf{b}}\cdot\mathbf{b}_i)-1)\right>$, where $\mathbf{b}_i$ is $i$-th bond between two particles, brackets are averaging across all bonds, and $\hat{\mathbf{b}}=\left<\mathbf{b}_i\right>$ is average bond direction. $P$ describes the orientating degree or a sample, with zero corresponding to fully amorphous samples and unit to fully oriented ones.

We also calculated entanglement length to describe how entanglement density affected the process of deformation. We used method Z \cite{kroger2005shortest} due to its computational efficiency needed to analyze large systems. This method uses geometrical criteria to determine whether chains represented as thin threads with kinks form an entanglement or not. The ends of chains were fixed in space and contour lengths were being minimized. At the end, the program calculated the average number of kinks in a chain $Z$. Entanglement length was calculated as $N_e = N/(Z+1)$. We used parameters $linethickness=0.00002$ and $threshold=0.02$ that differed from the default values to speed up the calculations. We also introduced the dependence of the neighbor list array on chain length and the number of chains in the system to avoid neighbor list array overflow \cite{petrov2021correction}.

Finally, we calculated 2D SAXS patterns. We removed all non-polymerized particles (solvent) and beads comprising the amorphous phase from the analyzed systems to reduce the noise level. Similarly to the experimental works, we calculated 2D SAXS patterns as if the incident rays were falling perpendicularly to the deformation axis (x-axis).
First, we constructed a 3D array consisting of elements calculated as in Eq. \ref{eq:scattering}.

\begin{equation}
\label{eq:scattering}
\begin{split}
    I_{kmp}=\frac{1}{8N}\left|\sum_{j=1}^{N}\sum_{t,f,w=1}^{2}\exp\left(2\pi i\left((-1)^tk\frac{x_j}{L_x}+(-1)^fm\frac{y_j}{L_y}+\right. \right. \right.\\
    \left. \left. \left. \vphantom {\sum_{j=1}^{N}}+(-1)^wp\frac{z_j}{L_z}\right)\right)\right|^2
\end{split}
\end{equation}

Here, $x_j$, $y_j$, and $z_j$ are the x-, y-, and z-coordinates of a j-th particle, respectively, and $k$, $m$, and $p$ are non-negative integers. The x-, y-, and z-components of the scattering vector are simply equal to $q_x=2\pi k/L_x$, $q_y=2\pi m/L_y$, and $q_z=2\pi p/L_z$. In Eq. \ref{eq:scattering}, we summed over the integers $t$, $f$, and $w$ to perform averaging over all conjugate complex numbers in order to obtain a clearer 2D SAXS pattern. This procedure does not lead to a loss of essential structural information due to rotational symmetry of the deformed polymer sample. Therefore, our 2D SAXS patterns will represent only the first quadrant of a usually circular pattern (since $q_x\geq0$ and $q_z\geq0$), which is sufficient for analyzing the microscopic mechanism of deformation \cite{tian2014lamellae,tian2015transition,jiang2010two}.

After this array had been calculated, we obtained 2D SAXS patterns. To gather more data and make 2D SAXS patterns clearer, we performed averaging as if the sample was rotated about the x-axis. The averaging procedure, however, gave rise to the following artifact on the 2D SAXS patterns: the values at $q_z=q_y=0$ remained non-averaged. Therefore, the values situated on the horizontal axis going through the center of the plots in Fig. \ref{fig:saxs} exhibited anomalous behavior. To build the plots in Fig. \ref{fig:saxs}, we replicated and mirrored the calculated quadrant and composed the replicas.

\section*{Supplementary Material}

\subsection{Fast Polymerization Procedure}

In this section, we describe a method of obtaining the $N_e50$ system. To obtain this system consisting of long chains with an equilibrium entanglement density, we implemented a specific fast polymerization routine. The other three studied systems were obtained following the procedures described in ref. \cite{Petrov20}.

First, we performed polymerization at $T=0.5$. All simulation parameters were taken as described in ref. \cite{Petrov20} with several exceptions. Namely, the probabilities of propagation and initialization were set to $p=1.0$, and the reactions occurred every $N_{stp}=10$ MD timesteps. The timestep was equal to $\Delta t=0.001$, the temperature and pressure damping parameters were equal to $1.0$ time unit ($\tau$). Therefore, the effective reaction rate defined as $r=p/(N_{stp}\Delta t)$ was equal to $r=100\tau^{-1}$ during the first stage of polymerization.

Second, after the conversion degree reached $f\approx 82\%$, we changed the temperature to $T=2.0$ and $N_{stp}$ to $N_{stp}=1.0$; as a result, the reaction rate increased to $r=1000 \tau^{-1}$. This allowed us to reach the target conversion degree of $f=90\%$ quicker; the system was also in a molten state, so we could perform equilibration as described in the next paragraph.

Third, we needed to equilibrate the system obtained after such fast artificial polymerization procedure. To do so, we set the temperature to $T=1.0$, and performed MD simulations without polymerization reaction for $9\times 10^7$ MD timesteps ($\Delta t=0.005$). The cutoff radii of 9-6 Lennard-Jones (LJ96) potentials acting between beads of all types were set to $R_{cut}=1.02$; hence, the synthesized polymer was equilibrated in athermal solvent conditions.

Finally, after the equilibration procedure, we carried out crystallization of the system. The polymer was gradually cooled to $T=0.5$ with constant rate $10^{-6}\tau^{-1}$. We changed the cutoff radius of LJ96 potential acting between non-polymerized particles (monomers) to $R_{cut}^{(MM)}=0.9$ before cooling, so the polymer was placed in a poor solvent at $T=0.5$ similarly to all other systems studied in our work. As a result, the semicrystalline material with $N_e=50$ and crystallinity $59\%$ was formed.

\begin{figure}[htbp]
\centering
  \includegraphics[width=\linewidth]{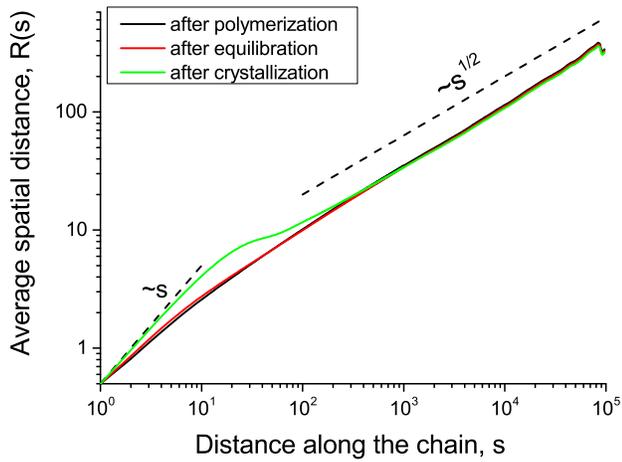}
  \caption{The $R(s)$ dependencies for the system after fast polymerization, after equilibration in athermal solvent, and after crystallization. Black dashed lines are included as a guide for the eye.}
  \label{fig:rs_ref_synth}
\end{figure}

The $R(s)$ dependencies characterizing conformation of chains after every step of the aforementioned procedure are shown in Fig. \ref{fig:rs_ref_synth}.

\subsection{Additional Figures}

\bibliography{rsc}
\bibliographystyle{rsc}

\section*{Acknowledgements}
We thank Pavel Kos and Alexey Gavrilov for illuminating discussions. The research was performed using the equipment of the shared research facilities of HPC computing resources at Lomonosov Moscow State University. The reported study was funded by RFBR according to the research project no. 18-03-01087. The research of Artem Petrov is supported partly by the grant no. 21-2-2-2-1 from the Foundation for the advancement of theoretical physics and mathematics “Basis”.

\section*{Author Contributions}
Artem Petrov: Conceptualization, Investigation, Methodology, Software, Writing - Original Draft. Vladimir Yu. Rudyak: Conceptualization, Investigation, Methodology, Software, Writing - Review \& Editing. Alexander Chertovich: Conceptualization, Supervision, Writing - Review \& Editing.

\section*{Competing Interests}
There are no conflicts of interest to declare.

\end{document}